\documentclass{aa}  

\usepackage{color,soul}
\usepackage{txfonts}
\usepackage{xcolor}
\usepackage{graphicx}
\usepackage{array}
\usepackage{amsmath}
\usepackage{float}
\usepackage{hyperref}
\usepackage{ulem}
\usepackage{makecell}
\usepackage{tikz}
\usepackage{subcaption} 
\usepackage{mathtools, nccmath}
\hypersetup{
    colorlinks,
    linkcolor={red!50!black},
    citecolor={blue!50!black},
    urlcolor={blue!80!black}
}
\makeatletter
\renewcommand*\aa@pageof{, page \thepage{} of \pageref*{LastPage}}
\makeatother
\bibpunct{(}{)}{;}{a}{}{,} % to follow the A&A style
\newcommand{\sj}[1]{}

\DeclareMathOperator*{\argmin}{\arg\!\min}
\newcommand{\remove}[1]{}
\newcommand{\removele}[1]{}
\newcommand{\nnew}[1]{#1}
\newcommand{\nnewle}[1]{#1}
\newcommand{\LEt}[1]{}
\begin{document}

   \title{Inverse-problem versus principal component analysis methods for angular differential imaging of circumstellar disks}
    \subtitle{The \texttt{mustard} algorithm}
    \titlerunning{Inverse-problem versus PCA methods for circumstellar disk imaging with ADI}
    
    \author{S.~Juillard\inst{1}\fnmsep\thanks{F.R.S.-FNRS FRIA grantee} 
          \and
          V.~Christiaens\inst{1}\fnmsep\thanks{F.R.S.-FNRS Postdoctoral Researcher}
          \and
          O.~Absil\inst{1}\fnmsep\thanks{F.R.S.-FNRS Senior Research Associate}}
      
    \institute{STAR Institute, Universit\'e de Li\`ege, All\'ee du Six Ao\^ut 19c, 4000 Li\`ege, Belgium \\ \email{sjuillard@uliege.be}}
    
    \date{Received 22 july 2023, Accepted 25 septempber 2023}

% \abstract{}{}{}{}{} 
% 5 {} token are mandatory
 \abstract
  % context heading (optional)
  {\LEt{ General notes: A.) I have edited to US English spelling and grammar conventions. B.) A\&A uses the past tense to describe specific methods used in a paper and the present tense to describe general methods as well as findings, including the findings of recent papers (within the past ten or so years). Kindly make any necessary changes (I have made some, but my edits are by no means exhaustive in this respect). See Sect. 6 of the language guide https://www.aanda.org/for-authors/language-editing/6-verb-tenses. ***}Circumstellar disk images have highlighted a wide variety of morphological features. Recovering disk images from high-contrast angular differential imaging (ADI) sequences is, however, generally affected by geometrical biases, leading to unreliable inferences of the morphology of extended disk features. Recently, two types of approaches have been proposed to recover more robust disk images from ADI sequences: iterative principal component analysis (I-PCA) and inverse problem (IP) approaches.} 
  % aims heading (mandatory)
  {We introduce \texttt{mustard}, a new IP-based algorithm specifically designed to address the problem of the flux invariant to rotation in ADI sequences -- a limitation inherent to the ADI observing strategy -- and discuss the advantages of IP approaches with respect to PCA-based algorithms.}
  % methods heading (mandatory)
  {The \texttt{mustard} model relies on the addition of morphological priors on the disk and speckle field to a standard IP approach to tackle rotation-invariant signals in circumstellar disk images. We compared the performance of \texttt{mustard}, I-PCA, and standard PCA on a sample of high-contrast imaging data sets acquired in different observing conditions, after injecting a variety of synthetic disk models at different contrast levels.}
  % results heading (mandatory)
   {\remove{As expected,} \texttt{Mustard} significantly improves the recovery of rotation-invariant signals in disk images, especially for data sets obtained in good observing conditions. However, the \texttt{mustard} model inadequately handles unstable ADI data sets and provides shallower detection limits than PCA-based approaches.}
  % conclusions heading (optional), leave it empty if necessary 
  { \texttt{Mustard} has the potential to deliver more robust disk images \nnewle{by introducing a prior}\removele{if a prior is introduced} to address the inherent ambiguity of ADI observations\LEt{ Verify that your intended meaning has not been changed. ***}. However, the effectiveness of the prior is partly hindered by our limited knowledge of the morphological and temporal properties of the stellar speckle halo. In light of this limitation, we suggest that the algorithm could be improved by enforcing a data-driven prior based on a library of reference stars.}


   \maketitle

%% ---------------------------- Introduction ------------------------------------------- %%

\section{Introduction}

High-contrast imaging (HCI) is a particularly challenging field as its primary goal is to enable the detection of objects near very bright stars, such as exoplanets and circumstellar disks that are typically \nnew{orders of magnitude fainter than their host star} \remove{$10^3$ to $10^{10}$ times fainter than their host star.} %\nnew{however, no disk fainter than $10^6~10^7$ has been detected so far}.\nnew{However, achieving these expected contrasts has not yet been realized.}
The combination of advanced instrumentation with a relevant association of observing strategies and post-processing techniques is required to reveal the faint circumstellar signal hiding in the vicinity of the star. The majority of the starlight and aberrations caused by atmospheric turbulence are removed by the coronagraph and adaptive optics, respectively. However, \nnew{in addition to the constraints of challenging observing conditions,} a significant remaining limitation lies in quasi-static speckles that are not mitigated by the adaptive optics system \citep{Faustine}. Observing strategies such as angular differential imaging \citep[ADI;][]{Marois06} \removele{enable }\nnewle{provide a lever for }\sj{We wished to emphasize how it enables differentiation by providing a lever}post-processing techniques to differentiate between quasi-static speckles and the circumstellar signal\LEt{ Verify that your intended meaning has not been changed. ***}. Angular differential imaging\LEt{ Avoid beginning a sentence with an acronym, abbreviation, number (unless written out), formula, or symbol. ***}  is an observing strategy in which the telescope pupil is stabilized with respect to the detector, and the field of view rotates over the course of the observation. The result is a sequence of images where the stellar halo and its associated speckles remain quasi-static over time, while circumstellar objects rotate with the field of view. Nowadays, ADI is the prime observing strategy used in most HCI surveys dedicated to searching for exoplanets and imaging circumstellar disks.
     
An important application of HCI techniques is the study of planet formation around young stars surrounded by protoplanetary disks, where planets are known to form. Recent observations \nnewle{with the Atacama Large Millimeter and submillimeter Array (ALMA)}\sj{Needed to be defined}\LEt{ Consider defining. ***} at submillimeter\LEt{ Units should be spelled out in full when not following a numeral. ***} wavelengths and with the latest generation of HCI instruments at near-IR wavelengths have revealed a wide diversity of structures in these disks. It is as yet unclear how all these structures are connected to embedded planets since very few protoplanets have been unambiguously confirmed to date \citep[e.g.,][]{PDS70info, Haffert19, Hammond23}. Principal component analysis (PCA) is currently one of the most used post-processing techniques for ADI sequences obtained for such systems. A common issue with PCA, and \nnew{many} state-of-the-art algorithms for ADI processing such as median-subtraction \LEt{ We reserve the use of slashes to denote ratios and instrument pairings and for use in equations. The use of "and/or" is acceptable. Kindly rephrase here and where needed. ***}\nnewle{and locally optimised combination of images} \LEt{ Consider defining. ***}\citep[namely Classical-ADI and LOCI; ][]{Marois06, Lafreni07}\sj{LOCI added in the citep}\nnew{, is that it \removele{was}\nnewle{is} designed to search for point-like sources and is not well suited for retrieving the geometry of extended disk signals. More specifically, ADI-based processing techniques have been shown to induce deformations on extended signals} \citep{Milli, valentin19, REXPACO, mayo}. These deformations compromise the interpretability of disks imaged using ADI and hence also studies of their morphological features, such as asymmetries or spirals, which are potential indirect planet signatures and could inform on ongoing planet--disk interactions. Moreover, these deformations can also lead to false-positive protoplanet candidates due to the filtering of extended signals into compact (point-like) sources \nnew{\citep[e.g., LkCa 15 b,c,d;][]{Currie19}}. %\nnew{Undoubtedly, disproving a protoplanet claim can be challenging, as numerous explanations can be found to account for certain inconsistencies. Consequently, confirmed false positives within the literature are rare, even though some of them have been subsequently invalidated through follow-up investigations. For instance, HD 100546 b, as originally presented by \citet[NACO-L']{Quanz2013}, has not received confirmation from subsequent SPHERE follow-up studies \citep{Sissa19}. Similarly, HD 163296 'b' \citep[NIRC2-L']{Quanz2013} remains unconfirmed, as \citet{Mesa19} did not provide supporting evidence.} 
 \nnew{Algorithms that leverage reference star differential imaging do not suffer from self-subtraction and associated geometrical deformations \citep[][]{Ren18, Ruane2019, Xie2022}.} \nnew{However, satisfactory reference stars are not available for all data sets. Therefore, the main goal of the present work is to accurately deduce the true morphology of a disk solely from ADI sequences. Under these assumptions, our prime objective is} to prevent any confusion between geometrical artifacts and real disk features, hence facilitating the interpretation of science images. In this context, some algorithms have been developed specifically for the imaging of extended sources in an attempt to prevent deformations. Two of them are based on inverse problem (IP) approaches \citep{REXPACO, mayo}, and two others are based on an iterative-PCA (I-PCA) algorithm \citep{mayo, StapperGinski22}. %\nnew{Alternatively, Non-negative Matrix Factorization \citep[NMF]{Ren18,Ren20} without data imputation is compatible with ADI only. sNMF with data imputation seems only feasible with an RDI strategy unless the presence of the disk on specific pixels, where it will rotate in the images, is known in advance.}

While \nnew{these methods} have been shown to provide a more robust recovery of disk images, they still suffer from the intrinsic limitation of ADI-based observations of not properly recovering rotation-invariant disk signals \nnew{\citep[also referred to as circular invariant flux in][]{Sandrine}}. Here, we present a new IP-based approach that aims to improve the recovery of rotation-invariant disk signals. In Sect.~\ref{sec:IPCA} we provide a general description of I-PCA algorithms and their limitations. Section~\ref{sec:IP} introduces our novel IP-based algorithm, \texttt{mustard}, and explains how it differs from previous IP-based implementations. Then, in Sect.~\ref{sec:eva}, we present a series of tests on ADI data sets with injected synthetic disks to compare the performance of IP and I-PCA approaches. In light of this analysis, we discuss in Sect.~\ref{sec:discussion} the advantages and drawbacks of IP approaches and provide recommendations as to which algorithm should be used to recover a disk from ADI sequences acquired in different conditions. 

In the rest of the manuscript, we use the following notations: $Y\in \mathbb{R}^{n,m,m}$ denotes the ADI sequence (also known as the ADI cube), with $n$ the number of frames, and where each frame is a square image of size $m\times m$ pixels. Each frame, $Y_k$, of the ADI data set is associated with a parallactic angle, $\theta_k$. We use $S\in \mathbb{R}^{n,m,m}$ to denote the cube of \nnew{speckle fields}\remove{stellar PSFs} reconstructed by a given algorithm, with $S_k\in \mathbb{R}^{m,m}$ the \nnew{speckle field}\remove{stellar PSF} in the $k^{\rm th}$ frame and $s\in \mathbb{R}^{m,m}$ the average \nnew{speckle field\LEt{ Verify that your intended meaning has not been changed. ***}}\remove{stellar PSF}. The 2D definite matrix $d\in \mathbb{R}^{+,m,m}$ denotes the circumstellar signal. The estimate of a parameter is written with an over bar ($\Bar{d}$). The PCA approximation of the ADI cube is denoted $\mathcal{H}_{q} : \mathbb{R}^{n,m,m} \longrightarrow \mathbb{R}^{n,m,m}$, where $q$ is the number of principal components (PCs) that are kept in the singular value decomposition. This operator provides an estimation of the \remove{cube}\nnew{speckle field}\remove{stellar PSFs}: $\mathcal{H}_{q}(Y) = \Bar{S}$. We use $\mathcal{R}_\theta(x)$ to denote the rotation operator for one frame, $x$, by an angle $\theta$, and $\Delta_{\theta}$ is a scalar denoting the total parallactic angle variation in a given data set. We use $\mathcal{P}(x)$\removele{for example}\nnewle{ to denote regularization fuctions} $\mathcal{P}:\mathbb{R}^{m,m}\longrightarrow\mathbb{R}$. \nnew{All vector and matrix multiplications} \remove{The matrix product $A \times B$}are based on the \nnew{element-wise} Hadamard product. The absolute value operator is written $||A||$. The $l_1$-norm $|A|_{l_1}$ is the sum of the absolute values of the elements in the \nnewle{vector or the matrix}\sj{Remove slash}\LEt{ slash. ***} $A$, while the $l_2$-norm $|A|_{l_2}$ is the sum of the absolute square values of the elements \removele{in the vector matrix}\sj{Clear without repetition, helps avoiding slash.}\LEt{ slash. ***}.
    
%% ---------------------------- Section inverse problem ------------------------------------------- %%

\section{I-PCA for ADI sequences}
\label{sec:IPCA}

\remove{PCA is currently one of the most used post-processing technique in HCI, despite being known for inducing ambiguous deformations of extended signals such as disks %\citep{valentin19, Milli, Sandrine}
.} Principal component analysis processing of an ADI sequence is based on the calculation of a PC basis through singular value decomposition of the ADI sequence, after the transformation of the image cube into a 2D (time vs. spatial) matrix. The PCA method creates a \remove{PSF model} \nnew{a model of the speckle field} by projection of the ADI images onto a chosen number of PCs. Then the model is subtracted from the original images, so that the residual images contain less stellar light while preserving circumstellar signals (captured in high-order PCs) as much as possible \citep{Soummer12, Amara12}. Even with a wise choice for the number of components, this method is systematically aggressive toward extended signals, which are always partly subtracted and hence not well restored. This is largely due to circumstellar signals appearing partly in low-order PCs \citep{Milli, valentin19}.
    
\nnew{An iterative process, involving the reestimation of the speckle field and the removal of previous estimations, was introduced by \citet{Milli} using classical ADI (cADI). They concluded that this approach could assist in mitigating self-subtraction. Subsequently, leveraging PCA methods,} I-PCA approaches were developed to iteratively refine the estimation of the \remove{stellar PSF}\nnew{speckle field} in each frame by iteratively removing the estimated disk signal to prevent it from being captured in the principal components (and hence in the subtracted \remove{PSF}\nnew{speckle field} model):
\begin{align*}
    \Bar{S}_{i+1} &= \mathcal{H}_{q}(Y - \mathcal{Q}(\Bar{d}_i)) \, , \\
    \Bar{d}_{i+1} &= ||\mathcal{Q}^{-1}(Y-\Bar{S}_{i+1})|| \, ,
\end{align*} 
where $\mathcal{Q} : \mathbb{R}^{m,m} \longrightarrow \mathbb{R}^{n, m,m}$ creates an image cube of $n$ frames, with each frame $k$ containing the same disk image $d$ rotated by the parallactic angle $\theta_k$. The inverse operation $\mathcal{Q}^{-1} : \mathbb{R}^{n,m,m} \longrightarrow \mathbb{R}^{m,m}$ corresponds to averaging the de-rotated cube of \remove{PSF}\nnew{speckle field}-subtracted frames. Throughout this paper, we use the \texttt{GreeDS} implementation of I-PCA described in \citet{mayo}. Our implementation of \texttt{GreeDS} was extracted from the \texttt{MAYONNAISE} package\footnote{\url{https://github.com/bpairet/mayo_hci/blob/master/GreeDS.py}} in order to be used independently. Our implementation is a re-factored version inspired from the original code, available on GitHub\footnote{\url{https://github.com/Sand-jrd/GreeDS}} and providing some extra options (see the GitHub page for details).

%, particularly in regions with negative areas.} 
Iterative PCA has been shown to provide better results in terms of disk image recovery than classical PCA \citep{mayo, StapperGinski22}, even though in their I-PCA tests on simulated data, \cite{StapperGinski22} have observed disk deformations that depend on the disk morphology and on the available degree of rotation. A common limitation to any PCA-based method when working with extended signals is the poor handling of the circularly symmetric component of the signal \citep{REXPACO}. This limitation is inherent to the ADI strategy, as circularly symmetric components are invariant to field rotation. For a given disk, $d$, and amplitude of rotation, $[\theta_0,\theta_n]$, we defined flux invariant to the rotation $f_{ir} \in \mathbb{R}^{m,m}$ as \remove{the ensemble of pixels}\nnew{the matrix containing the minimum flux in each pixel for all considered parallactic angles,} $\theta_k$. \remove{The amplitude of the ambiguous flux corresponds to the minimum of the intensity that a pixel can take while the disk rotates:}\nnew{This matrix can be expressed as follows:}
\begin{equation}
    f_{ir} : \forall \ x,y \in [0,m] \, ,  f_{ir}(x,y) = \min_{\theta \in [\theta_0..\theta_n]} \mathcal{R}_{\theta}(d)(x,y) \, .  
\end{equation}
State-of-the-art HCI algorithms suffer from a lack of a circularly invariant component in their output disk estimate due to their incapacity of distinguishing the part of the disk flux invariant to the rotation from a static component (i.e., speckle field).
%In most data sets, both disk and speckle field morphologies contain flux invariant to the rotation. Depending on how aggressively the algorithm will remove the quasi-static contribution from the data set, more deformation will appear.

\begin{figure}[!t]
    \centering
    \includegraphics[width=\linewidth]{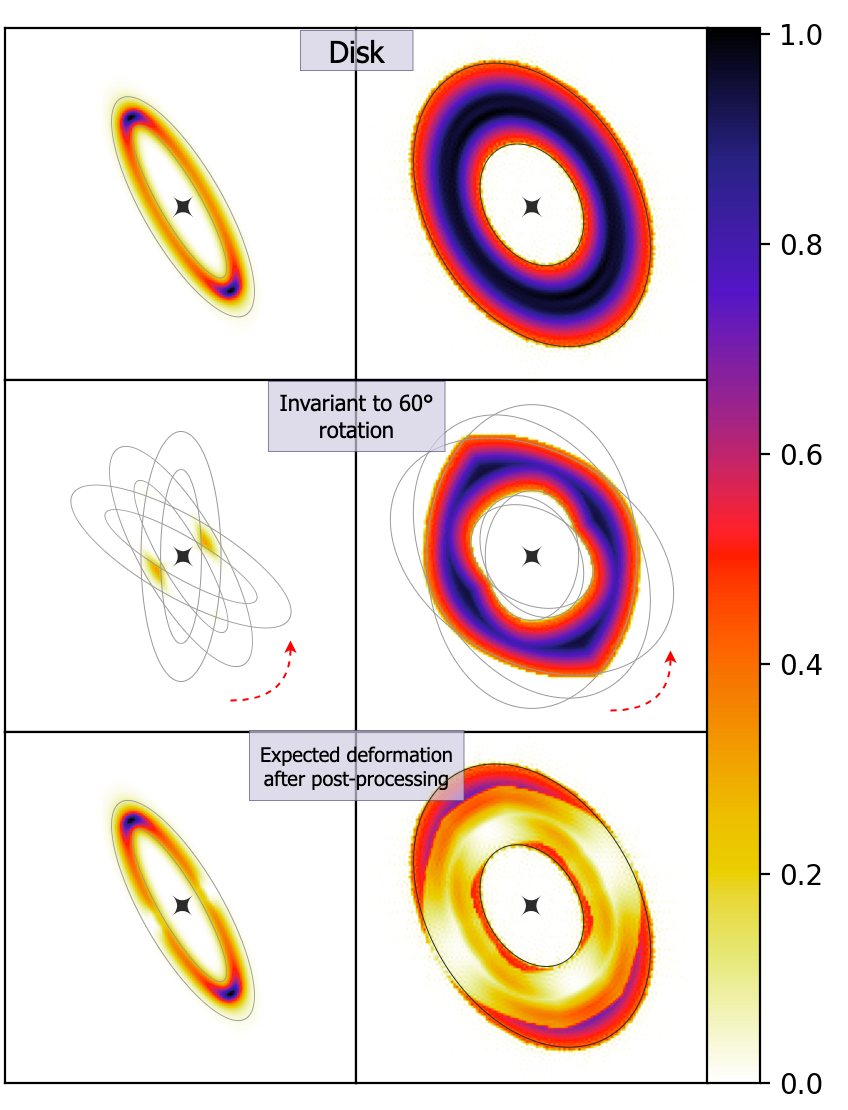}
    \caption{Illustration of flux invariant to the rotation in disk images, showing two disk geometries (top), their region invariant to the rotation (middle), and the expected deformation induced by post-processing leveraging ADI (bottom).\ White areas are expected to be self-subtracted by post-processing. Intersections are computed with Python using ten frames\LEt{ Write out whole numbers when lower than 11 (i.e., zero to ten) and not directly used as a measurement with the unit following; numbers 11 and up should be written numerically (unless at the beginning of a sentence). See Sect 2.7 of the language guide https://www.aanda.org/for-authors/language-editing/2-main-guidelines. ***} evenly rotated between 0 and 60 degrees}
    \label{fig:pyambig}
\end{figure}

A couple of examples of flux invariant to the rotation for different disk morphologies are shown in Fig.~\ref{fig:pyambig} for an amplitude of rotation $\Delta_\theta = 60^{\circ}$. The deformations associated with rotation-invariant flux do not only appear in face-on disks, but are a systematic problem that appear in all disk morphologies. The rotation-invariant regions are more significant for wider disks, low inclination, and low amplitude of rotation. Leveraging the angular rotation only is not enough to correct for these deformations. Indeed, these regions are common to all frames, and thus share the same characteristic as the reconstructed speckles field. It is not always obvious to guess the initial disk morphology from the ADI post-processed images, considering the wide variety of geometries and substructures a disk can have. This limitation is inherent to the observing strategy, and can only be corrected by introducing additional information.

%*******************************************

\section{Inverse problem approaches for ADI sequences}
\label{sec:IP}

An IP approach is a statistical method that aims to estimate causal factors from the observed effects they produce. %\citep{PBinverse}.
A general expression of the method can be defined by considering a set of observations ($Y$) corrupted by an error $\epsilon$. The observations are the result of a known causal process (\ref{forward}, $\mathcal{F}$) for a given set of input parameters ($x$).  The residuals between observations and estimations of the data $\mathcal{F}(\Bar{x})$ are called the misfit. Ideally, if the model is a good description of the reality, the misfit should be equal to the noise $\epsilon$. It is required to use an estimator to invert the process (\ref{backward}),  in order to find the optimal estimation $\Bar{x}$ that will minimize the misfit. The choice of the estimator is determined by the assumptions on the statistical distribution of the error $\epsilon$. In the case of mean square error (MSE), $\epsilon$ is considered to be white noise, and the IP is described as follows:
\begin{align} 
        Y &= \mathcal{F}(x) + \epsilon \, ,\tag{Forward model}\label{forward}\\
        \Bar{x} &= \argmin_{x}{\bigr[|Y - \mathcal{F}(x)|_{l_2}\bigr]} \, . \tag{Backward estimator}\label{backward}
\end{align}
According to the Hadamard conditions \citep{invProb}, a well-posed problem requires that a solution (i) must exist, (ii) is unique, and (iii) changes continuously with the initial conditions. In most applications, the problem to be solved is, however, ill posed, which means that at least one of the Hadamard conditions is not fulfilled.

The key for an IP approach to tackle ill-posed problems is regularization (or constraint), which is the process of adding an extra source of information in order to circumvent the inherent ambiguities of the model, and transform the ill-posed problem into a well-posed problem. This practice, when used with temperance, can be a great tool for tackling difficult problems. The downside, however, is that artificially good results can easily be obtained based on unsubstantiated priors that do not correctly answer the addressed scientific questions. With the addition of regularization, an IP algorithm relies on the minimization of two terms: first, the data-attachment term given by the estimator and second, the regularization term(s) that aims to constrain the solution space with prior knowledge on $x$, leading to the following backward estimator:
\begin{equation*} 
    \Bar{x} = \argmin_{x}{\bigr[|Y - \mathcal{F}(x)|_{l_2} + \mu \mathcal{P}(x)\bigr]} \, .
\end{equation*}
When applying an IP approach, it is crucial to understand the scope of validity of the chosen forward model and its inherent ambiguities, to be aware of the validity of the prior added through regularization, and to wisely adjust its strength to guarantee robustness and trustworthiness of results. This is handled by the hyperparameter $\mu$, whose goal is to define the regularization strength.  The value of $\mu$ generally needs to be adjusted by the user to strike an appropriate balance between the data attachment term and the regularization.
Overall, an IP algorithm is a flexible method that can be adapted in various manners by changing the estimator and regularization. One purpose of using an IP approach over an estimator like PCA is to provide a better interpretation of the data through a model, and by adding the lacking information through regularization.

Applied to ADI, the scientific task we wish to tackle with an IP algorithm is to reliably sort between \remove{stellar halo}\nnew{speckle field} and the azimuthally extended (circularly invariant) disk signal. However, ADI data sets are a very adversarial environment. We can sort the sources of errors into three categories: (i) inherent ambiguities of angular diversity (flux invariant to the rotation, small amplitude of rotation, uneven distribution of parallactic angles), (ii) data out of the scope of validity (fast varying speckles, wind-driven halo, undetermined noise statistics), and (iii) technical bottlenecks. The latter may include algorithmic elements, such as the limitations of the rotation operator in the discrete domain, or issues related to a specific method, such as non-convexity for gradient descent.

\subsection{The \texttt{mustard} approach}
\label{sec:must}

Our motivation to propose a new IP algorithm for disk imaging in HCI is to focus on the inherent ambiguities of the ADI technique, as we noticed that no algorithm so far addresses this issue. We aim to correct for the geometrical biases of standard point-spread-function-subtraction techniques, which generally assign any rotation-invariant flux fully to the reconstructed \remove{PSF}\nnew{speckle field}, by making an assumption on its \remove{the} morphology \remove{of the speckle field}.

In order to build a statistically viable IP backward model to estimate the speckle field, it is required to assume the speckle field as static despite its stochastic behavior. The forward model $\mathcal{F}(s,d,k)$ that describes a frame $Y_k$ of the ADI cube is written as
\begin{equation*} 
    \mathcal{F}(s,d,k) = s + \mathcal{R}_{\theta_k}(d) + \epsilon_k \, .
    \label{equ:y_k}
\end{equation*}
The MSE estimator to invert this model is thus
\begin{equation} 
    \Bar{d},\Bar{s} = \argmin_{s\in \mathbb{R}^{+ m,m}, \, d\in \mathbb{R}^{+ m,m}}\biggr[\sum^{m\times m}_{pixels=0}\sum^{n}_{k=0} ||Y_k - (s + \mathcal{R}_{\theta_k}(d))||^{2}\biggr] \, .
    \label{equ:crit}
\end{equation}
% Considering a stable data set (i.e., a data set where the speckle pattern is close to a strictly static behavior, with no  and no strong variation of intensity between frames)
By construction, modeling the speckle field as static is equivalent to estimating the average speckle field $S\in \mathbb{R}^{n,m,m}$ as $s\in \mathbb{R}^{m,m}$. Ideally, variations in the speckle field will be attributed to the misfit, but this is not guaranteed.

Our IP is ill posed due to the overlap between $d$ and $s$. Indeed, a component invariant to rotation can be labeled as either static (i.e., as $s$) or rotating (i.e., as $d$). To sort out these ambiguities, we \removele{have}\nnewle{had} to use prior knowledge. For this purpose, we added a prior to the morphology of the \nnew{speckle field %to capture the flux invariant with respect to rotation originating from the stellar residuals. We define \nnew{
through the definition of a regularization mask, $M_{\mathrm{reg}}$, as a double-Gaussian profile %matching 
representing the expected amount of rotation-invariant flux associated with the speckle field. This mask was used to define a 
regularization term $\mathcal{P}_{\rm mask}$, which will guide the flux invariant to rotation of $\Bar{S}$, defined as follows:}

\begin{equation*}
    \centering
    \mathcal{P}_{\rm mask}(d) = |M_{\rm reg}\times d|_{l_1} \, ,
    \label{equ:Rl1}
\end{equation*}
where the linear norm, $l_1$, is weighted by the mask, $M_{\rm reg}$. 

\begin{figure}[!ht]
    \centering
    \includegraphics[width=\linewidth]{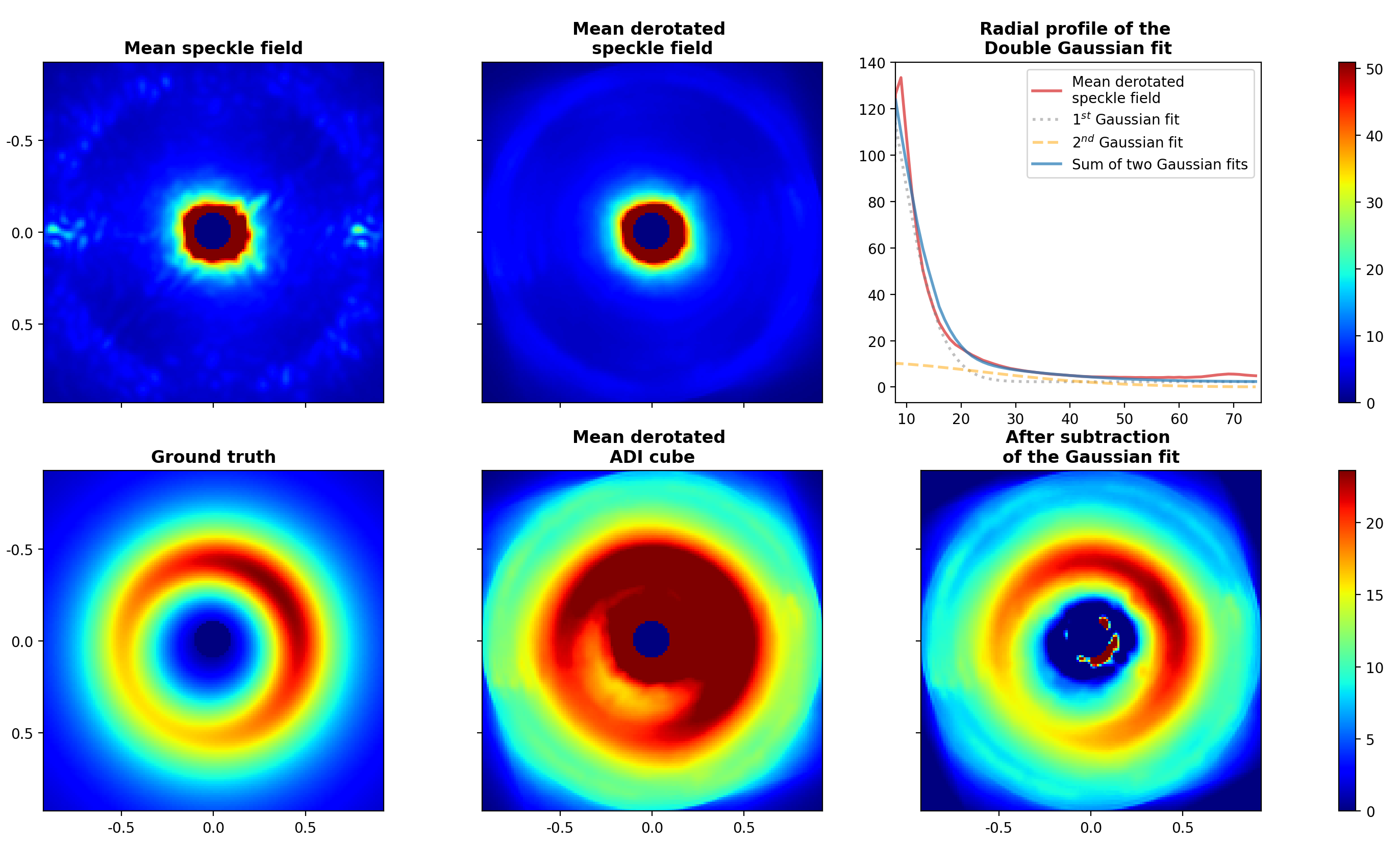}
    \caption{\nnew{Illustration of the capabilities of an optimal double-Gaussian mask to capture the rotation-invariant stellar residuals. The leftmost two figures in the first row respectively display the mean and the mean de-rotated speckle field %\remove{and the stellar halo}
    before injection of the disk signal. The top-right figure displays the radial profile of the de-rotated empty data set (solid red line) together with the double-Gaussian fit, with the two Gaussians represented in dotted gray and dashed orange, respectively. The solid blue line represents the sum of the two Gaussians. The fit was performed while excluding the pixels within the numerical mask, which is an aperture with an 8-pixel radius. %The estimated parameters include the coordinates of the centroid, FWHM in both the x and y directions, and the amplitude. 
    The bottom row shows, from left to right, the disk GT and the mean de-rotated cube after injection before and after subtraction of the optimal Gaussian mask. The figure scales are consistent across all images of each row. Axis values are in arcseconds. The intensity values displayed by the colorbar are arbitrary. This test was conducted using cube \#2 with disk D injected at a contrast of $5\times10^{-4}$.}}
    \label{fig:mask_subpot}
\end{figure}

\nnew{In Fig.~\ref{fig:mask_subpot} we show the anticipated impact of using a mask made up of a double 2D Gaussian. The first Gaussian (dotted gray in the top-right panel) is meant to represent the inner halo primarily arising from stellar residuals near the center, while the second Gaussian (dashed orange) is intended to capture the spread of the diffraction pattern across the entire frame. For illustration purposes, the two 2D Gaussian models are derived by a fit to the mean de-rotated cube image (top middle panel) before introducing the disk signal, and then subtracted from the mean de-rotated cube after the disk signal has been injected (bottom-right panel). This example is meant to demonstrate the impact of subtracting an optimal mask, which would help resolve the circular ambiguities, even though it does not exactly illustrate the process undertaken by the algorithm when such a prior is added as a regularizer. Obtaining an optimal Gaussian mask using actual data is challenging since do not have access to an empty data set. In practice, when using \texttt{mustard}, the parameters of the double Gaussian are thus set manually, which limits the efficacy of this approach in real observations, highlighting the critical importance of mask selection. 
Nonetheless, by utilizing the mask as a form of regularization rather than relying on it solely for fitting (i.e., incorporating it into the forward model by decomposing the static speckles and the diffraction pattern), we mitigated potential inaccuracies that may arise from imperfect mask definition.} %In fact, fitting a Gaussian as depicted in Fig.~\ref{fig:mask_subpot} is challenging and demands tight constraints on its parameter bounds to prevent errors (i.e., fitting one artifact instead of the entire stellar halo). This is why a single Gaussian is presented, as opposed to a double Gaussian corresponding to the definition of $M_{\rm reg}$ in MUSTARD. 
\nnew{In practice, the mask can be adjusted manually when using \texttt{mustard} in an attempt to minimize the distance between the model and the data. %As such, the double Gaussian model mask $M_{\rm reg}$  provides a prior of the speckle field's flux invariant to the rotation. 
This regularization} will be counterbalanced by the smoothness regularization and the data attachment term, both of which capitalize on angular diversity. Furthermore, the Gaussian approximation is suboptimal, as it fails to precisely describe the mean speckle field, including the bright circular halo around the region that is well corrected by the deformable mirror, at a radius of $0\farcs70$ in the data shown here.

\nnew{The parameters of the double-Gaussian mask have to be defined by the user when using \texttt{mustard}.} % While fine-tuning of the mask from a data set to another is expected to improve our results, we rather considered a realistic case scenario of not \remove{knowing} \nnew{having access to the ground-truth's} \remove{the stellar halo}\nnew{speckle field} profile \nnew{(due to the potential presence of a disk). Hence, unlike what was demonstrated in Fig~\ref{fig:mask_subpot}, we preferred not to utilize the information from the empty dataset before disk injection, }\remove{due to the potential presence of bright disk signals}.
Utilizing a more complex mask (e.g., learned from a library of \remove{PSFs}\nnew{coronagraphic observations of stars}) could improve our results but goes beyond the scope of this study, which solely aims to leverage ADI, and will be investigated in a future work. \texttt{Mustard} is initialized with the mean de-rotated ADI sequence (as in a no-ADI data set) in order to assign all ambiguous flux to the disk as a first estimate. Hence, this regularization will push out the ambiguous flux that does not belong to the disk according to the shape of the mask. The mask regularization $\mathcal{P}_{\rm mask}$ is defined as follows:
%An example of mask is given in Fig.~\ref{fig:exemple_mask}. 

% \begin{figure}[t]
%     \centering
%     \includegraphics[width=\linewidth]{Images/typical_mask.png}
%     \caption{Example of a mask used for regularization}
%     \label{fig:exemple_mask}
% \end{figure}
In addition, we defined a smoothness regularization term using the spatial gradient of the circumstellar and stellar signals $d$ and $s$, respectively. It is a common regularization to compensate for noise \nnew{\citep{IPregul}}. With $\Delta$ standing for the Laplacian operator, the two regularization terms read
\begin{align*}
    \mathcal{P}_{\rm smooth}(d) = ||\Delta d||^2 \, , \\
    \mathcal{P}_{\rm smooth}(s) = ||\Delta s||^2 \, .
\end{align*}

The three regularization terms are added to the data-attachment term, weighted by hyperparameters $[\mu_1,\mu_2,\mu_3]$. Additionally, a binary mask $M_c$ is added to ignore pixels located under the coronagraphic mask, or too close to the optical axis. The final expression for the minimization reads as follows: 
\begin{align}
    \Bar{d},\Bar{s} &= \argmin_{s\in \mathbb{R}^{+,m,m}, d\in \mathbb{R}^{+,m,m}}\biggr[\sum_{pixels=0}^{m\times m} \biggr(M_c \sum_{k=0}^{n} ||Y_k - (s + \mathcal{R}_{\theta_k}(d))||^{2}\biggr)\nonumber \\
    & + \mu_1\mathcal{P}_{\rm mask}(d) + \mu_2\mathcal{P}_{\rm smooth}(d) + \mu_3\mathcal{P}_{\rm smooth}(s)\biggr] \, .
    \label{equ:critReg}
\end{align}
A good balance between the morphological regularization terms and the data-attachment term is required to provide good results.
Hyperparameters can be set based on the initialization, $[d_0, s_0],$ of the disk and average speckle field, as a given percentage, $p_i$, of the data-attachment term:
\begin{equation}
\mu_i = p_i\frac{|Y - (s_0 + \mathcal{R}(d_0))|_{l_2}}{\mathcal{P}_{i,0}} \, ,
\label{equ:mu}
\end{equation}
where $\mathcal{P}_{i,0}$ is the value of the regularization term $i$ at initialization, namely  $\mathcal{P}_{\rm smooth}(d_0)$, $\mathcal{P}_{\rm smooth}(s_0)$, and $\mathcal{P}_{\rm mask}(d_0),$ respectively. \LEt{ We do not allow the use of "e.g." or "i.e." within the main text (in parentheses or within figure/table captions is fine). ***}For the initialization of the disk $d_0$, we used the positive mean de-rotated cube, while $s_0$ was set as the mean positive residuals after subtracting the disk initialization from the cube: $s_0 = \mathrm{max}\{0, \frac{1}{n} \sum_{k=0}^{n} Y_k - \mathcal{R}_{\theta_k}(d_0)\}$.

\texttt{Mustard} is a Python object-oriented software package using PyTorch and its implementation of the Broyden-Fletcher-Goldfarb-Shanno algorithm \citep{BFGS}, an iterative method for solving unconstrained nonlinear optimization problems. The Fourier-transform based image rotation used in \texttt{mustard} is a Torch tensor adaptation of its implementation in the VIP package\footnote{\url{https://github.com/vortex-exoplanet/VIP}}. The \texttt{mustard} package is available on GitHub\footnote{\url{https://github.com/Sand-jrd/mustard}}.

\subsection{Previous IP approaches: \texttt{MAYO} and \texttt{REXPACO}}
\label{sec:ip-comp}

\bgroup
\begin{table*}[t]
\caption{Comparison of assumptions made in the PCA, I-PCA, \texttt{REXPACO}, \texttt{MAYO}, and \texttt{mustard} algorithms with respect to reality.}
\def\arraystretch{1.5}
\begin{tabular}{>{\raggedright}p{0.10\linewidth} >{\raggedright}p{0.11\linewidth} >{\raggedright}p{0.12\linewidth} >{\raggedright}p{0.125\linewidth} >{\raggedright}p{0.125\linewidth} >{\raggedright}p{0.125\linewidth} >{\raggedright}p{0.13\linewidth}}
\hline\hline
\textbf{} &
  \multicolumn{1}{l}{PCA} &
  \multicolumn{1}{l}{I-PCA} &
  \multicolumn{1}{l}{\texttt{REXPACO}} &
  \multicolumn{1}{l}{\texttt{MAYO}} &
  \multicolumn{1}{l}{\texttt{mustard}} &
  \multicolumn{1}{l}{Reality} \\ \hline

Speckle field &
  PCs\tablefootmark{a} &
  Positive, PCs &
  Positive, static &
  Positive, PCs from I-PCA &
  Positive, static &
  Positive, quasi-static \tabularnewline
  
Disk signal &
  Rotating &
  Positive, rotating &
  Positive, rotating &
  Positive, rotating, shearlets &
  Positive, rotating &
  Positive, rotating \tabularnewline
  
Noise &
  N/A &
  N/A &
  Learned statistics, by patches, from background &
  Constrained\tablefootmark{b}, white noise  &
  White noise (implicit)\tablefootmark{c}   &
  White noise, shot noise, Rician speckle noise, + others\tablefootmark{d}\tabularnewline
  
Flux invariant to rotation &
  All belong to speckle field &
  All belong to speckle field &
  All belong to speckle field &
  All belong to speckle field &
  Sorted according to Gaussian mask\tablefootmark{e} &
  A fraction belongs to the disk\tabularnewline
\hline
\end{tabular}
\tablefoot{
\tablefoottext{a}{Principal component(s). Using one PC means that the estimated speckle field can vary in amplitude, while more components will capture variation in its morphology.}
\tablefoottext{b}{Noise statistics learned from I-PCA results and enforced within Huber loss.}
\tablefoottext{c}{Scope of validity of MSE estimator.}
\tablefoottext{d}{Other nuisance phenomena, which create temporally varying extended structures, such as wind-driven halos \citep{Faustine}, can be accounted for noise.}
\tablefoottext{e}{$M_{\rm reg}$ drives the rotation-invariant flux assignment (see Sect.~\ref{sec:must}).}
}

\label{table:algo}
\end{table*}
\egroup
\texttt{REXPACO} \citep[for Reconstruction of EXtended features by
PAtch COvariances;][]{REXPACO} is an IP approach that includes a statistical model of noise inherited from the \texttt{PACO} algorithm \citep{PACO}. 
\texttt{REXPACO} and \texttt{mustard} are similar regarding the data-attachment term, but there are two main differences between these algorithms. Firstly, one of the regularization terms of \texttt{REXPACO} penalizes the $l_1$-norm of the disk estimate, while in \texttt{mustard} we used a mask that aims to prevent flux ambiguous to the rotation from being systematically assigned to the speckle field estimate. Secondly, \texttt{REXPACO} \remove{influences}\nnew{enforces} the statistics of the misfit by using the covariance patches approach from the \texttt{PACO} algorithm.

\texttt{MAYONNAISE} \citep[or \texttt{MAYO} for short,][for %morphological components analysis pipeline for circumstellar discs and exoplanets imaging in the near-infrared
Morphological Analysis Yielding separated Objects iN Near infrAred usIng Sources Estimation]{mayo}, is an optimization algorithm that introduces a morphological analysis of the disk and planets after the speckle field estimation of the I-PCA algorithm, \texttt{GreeDS}. It differs in the way it addresses the problem: contrary to \texttt{REXPACO} and \texttt{mustard}, \texttt{MAYO} estimates a different speckle field for each frame. To not have the speckle field described as an unknown stochastic phenomenon, \texttt{MAYO} forces the speckle field $S$ to be described as a linear combination of the $q$ first PCs estimated through I-PCA. One of the main goals of \texttt{MAYO} is to deconvolve the circumstellar signal, and to separate point-like sources from extended disk signals.
% Which mean that for each frame $k\in [0,..n]$.
% \begin{equation*} 
%     Y_k = S_k + \mathcal{R}_{\theta_k}{d}
% \end{equation*}
% \begin{equation} 
%     \Bar{d} = \argmin_{S \in \mathcal{P}^{n,M,M}, d\in \mathbb{R}^{+,M,M}}{(\sum_{k=0}^{N}||Y_k - S_k - \mathcal{R}_{\theta_k}{d}||^2)}
%     \label{eq:estimator1}
% \end{equation}
The model therefore contains the convolution by the point spread function, and the disk and planet models are constrained by morphological priors using shearlets and point-like sources, respectively. Additionally, it uses a Huber loss estimator (instead of MSE) because this distance is strongly convex at the neighborhood of 0 and affine at its boundary, and hence less sensitive to outliers in the data than the squared error loss.

%\subsection{Theoretical comparison between different IP approaches}

To understand the specifics of each algorithm, and what kind of answer they can provide, it is necessary to comprehend the underlying assumptions made by the equations used in each model, and their dependences. 
Estimators that are grounded in physics and rely on well-established probabilistic methods are generally more convincing. They are considered more interpretable and reliable, although this \remove{is} may vary depending on the specific context and application \citep{EstimationTheory}.
The assumptions made by each algorithm are summarized in Table~\ref{table:algo}, and further discussed in the following bullet points, along with  their estimator properties.

\begin{table*}[t]
    \caption{Data sets used in our study obtained in the $\bold{H2}$ band with the IRDIS camera of VLT/SPHERE within the SHINE-F150 survey \citep{Desidera21,Langlois21}. \nnew{Both seeing and coherence time measurements come from the MASS-DIMM sensor}. These data sets are said to be ``empty,'' in the sense that they do not contain any known circumstellar disk or low-mass companion.}
    \centering
    \begin{tabular}{ c c c c c c c c c c}
    \hline\hline
        ID & Star & Date & \shortstack{Rotation \\(°)}&  \shortstack{Seeing \\ ('')} & \shortstack{Wind \\ speed (m/s)} & \shortstack{Strehl \\ratio} & \shortstack{Coherence \\time (s)} & \shortstack{Nb of \\ frames} & \shortstack{Comment} \\ \hline
        1 & HIP 107345 & 2015-07-04 & 26 & 1.07 & 11 & 0.49 & 2.1 & 64 & \shortstack{Low Strehl ratio}\\ 
        2 & HIP 77457 & 2015-03-30 & 115 & 0.94 & 17 & 0.79 & 1.7 & 60 & \shortstack{Wind-driven halo}\\ 
        3 & \shortstack{LP 776-25} & 2016-01-17 & 91 & 0.45 & 6 & 0.79 & 2.9 & 72 & Unstable \\ 
        4 & HIP 103460 & 2016-09-17 & 80 & 0.88 & 9 & 0.87 & 4.3 & 64 & Stable\\ 
        \hline
    \end{tabular}
    \label{tab:datasets}
\end{table*}

% This table does not capture the dependency between assumptions, as one assumption's validity can neutralize the others. In fact, since the model of the data is not a perfect representation of reality, the error, or misfit is no longer equal to the noise as stated in the equations Section.~\ref{sec:IP}.
\begin{itemize}
    \item The PCA model treats the speckle field as the PCs of \nnewle{the singular value decomposition (SVD)}\sj{Added definition}\LEt{Please define. ***}. This is an efficient method for removing the quasi-static part of the stellar halo, but it is too aggressive and can remove a significant part of the circumstellar signal \citep[self-substraction,][]{Milli,valentin19,StapperGinski22}. Additionally, it does not introduce any physical assumptions, such as the positivity of the disk signal, which could potentially lead to more accurate results. 
    \item The \remove{\texttt{GreeDS}}\nnew{I-PCA} algorithm\remove{(and I-PCA methods in general)} provides a PCA-based estimation of the disk while also enforcing the positivity of the disk signal. This approach corrects PCA self-subtraction effectively, but it does not correct for flux invariant to rotation.
    \item \texttt{MAYO} implements a morphological analysis by modeling the circumstellar signal as the sum of shearlet-based components and point-like sources. This model provides a prior for the deconvolution task, leading to \remove{sparse and} visually appealing outputs resembling disks \nnew{(i.e., enforcing smoothness, sharpness, or sparsity without statistical justification)}. However, the assumption of a unique speckle field for each frame is too unconstrained, which hinders the algorithm's ability to improve upon the initial estimate provided by \remove{\texttt{GreeDS}}\nnew{the I-PCA}. Moreover, a morphological analysis alone may not be sufficient to differentiate between signals and artifacts, as there is no distinct morphological component that characterizes them unambiguously. Therefore, it becomes necessary to limit the number of shearlets and planets, which could lead to an erasure of the compact signals, fine structures in the disks, or planets. Overall, the \texttt{MAYO} estimator suffers from identifiability issues: multiple sets of parameter values can fit the data equally well when the model is unconstrained, potentially resulting in ambiguous or unreliable estimation results  \citep{Identifiability&estimability}.
    The algorithm is prone to inheriting imperfections from the initial speckle field estimation process, so that its performance depends on the choice of rank and iteration parameters \nnew{used} in \remove{\texttt{GreeDS}}\nnew{the I-PCA}.
    \item The \texttt{REXPACO} algorithm uses an IP approach that introduces an estimation of the noise statistics and enforces the positivity of the signal. Along with \texttt{mustard}, the two algorithms do not use a PCA estimation to model the speckle field. However, to do so, these algorithms must assume that the speckle field is static. These estimators suffer from adequacy issues: the optimality of the probabilistic-based estimator is no longer guaranteed due to the model's failure to adequately represent the underlying system \citep{Inadequacy}. If the static assumption made on the speckle field is not fulfilled, then the underlying assumption that the misfit is white noise is invalid, and some speckle field variations will be fit ``at best'' into the rotating components (disk and planets) estimates. This can be the case for data sets acquired in challenging observing conditions or, more generally, for disks that are significantly fainter than the amplitude of speckle variations. In such cases, PCA-based estimations will better handle the variation in flux and speckle morphology between frames, by projecting each frame onto a well-chosen number of PCs.
    \item So far, \texttt{mustard} is the only approach that attempts to address the limitation of flux invariant to rotation, but it does not provide strong guarantees of robust disk image extraction as the parameters (width and amplitude) of the Gaussian mask have to be assumed a priori.
\end{itemize}

Finally, one of the main limitations to the recovery of extended sources from ADI sequences is the wind-driven halo. This type of nuisance elongates the stellar halo in a specific direction, creating a butterfly pattern \citep{Faustine}. Its structure varies throughout the acquisition, and rotates in a similar way as the field during ADI sequences. Despite being very common, this phenomenon goes beyond the scope of validity of all IP algorithms to date. 

% However, it is limited by our imprecise knowledge of the speckle field, and the arbitrary choice of mask $M_{\rm reg}$ and regularization weights $\mu$ may significantly alter the results.
% This belong to discussion : Optimal parameterization is a known limitation of the IP approach in general, and it is deterministic for both \texttt{mustard} and \texttt{MAYO}, as they will modulate the interpretation produced: \texttt{mustard} needs a mask $M_{reg}$ to describe the flux invariant to rotation, and \texttt{MAYO} needs to define the parameters for \texttt{GreeDS} first estimate, the number of planets, and the number of shearlets used to describe the disk. 

%% ---------------------------- Section  ------------------------------------------- %%

\section{Performance comparison}
\label{sec:eva}

In this section we conduct a comprehensive study to test the performance of \texttt{mustard} on a range of synthetic disks, and compare it with results obtained with PCA and I-PCA methods. We leave a complete performance comparison with other IP-based algorithms to a future disk imaging data challenge, where the developers of each method will have a chance to fine-tune their algorithm and provide the fairest possible comparison (more about this can be found in Sect.~\ref{sec:comparison}). We still provide in Sect.~\ref{sec:realdata} a couple of examples of \texttt{mustard} results on data sets that have been previously processed by \texttt{MAYO} and \texttt{REXPACO} in the literature.

\begin{figure*}[t]
    \centering
    \includegraphics[width=\linewidth]{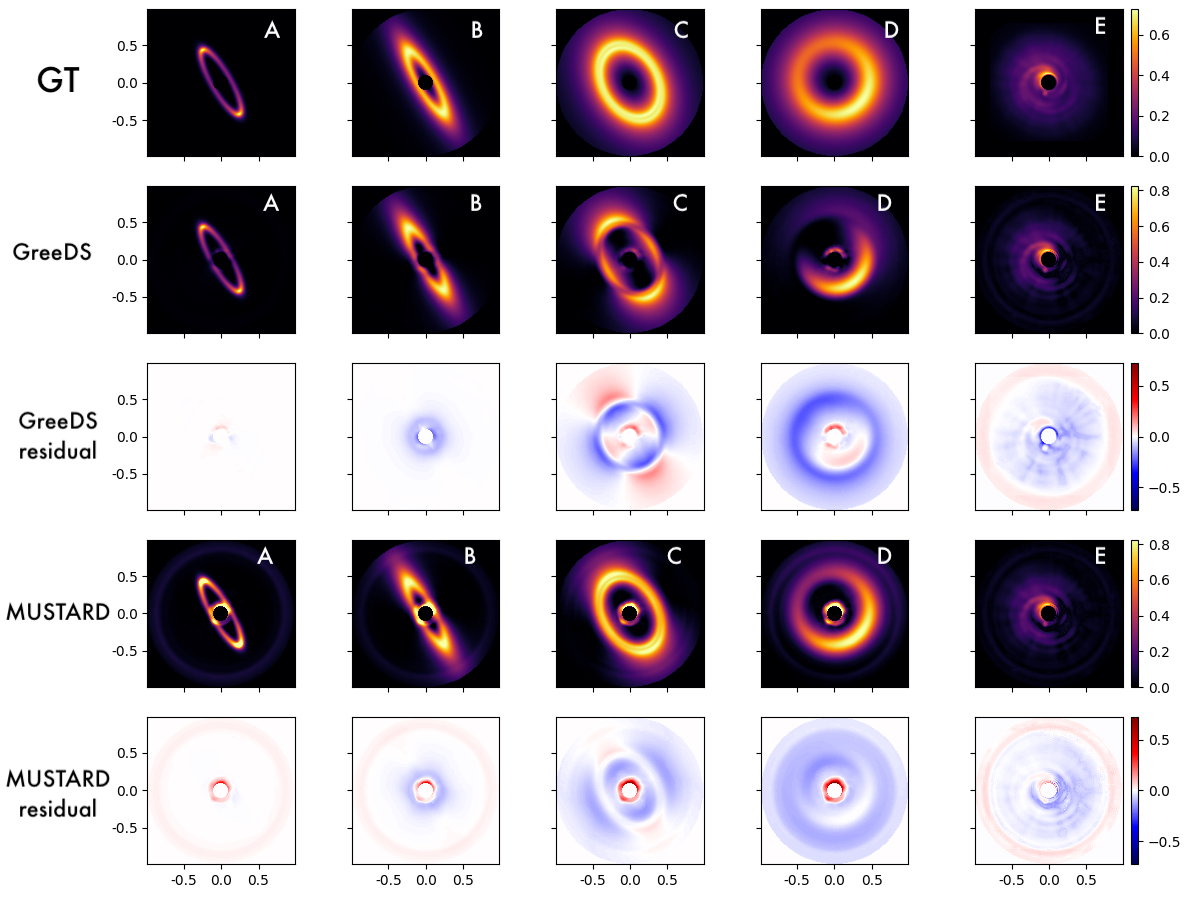}
    \caption{Disk estimations for a contrast of $10^{-3}$ injected in the ADI cube n°4 (see Table~\ref{tab:datasets}). Top: GT (injected disk images). Second and third rows: \nnew{I-PCA estimation using \texttt{GreeDS} implementation} and its residuals. Last two rows: \texttt{mustard} estimation and its residuals. In the residual maps, we emphasize the missing disk signal in blue, and excess signal in red. To highlight morphological biases only (i.e., not absolute flux recovery), disk estimations were rescaled in flux before the computation \nnew{of the difference with the ground truth} (GT) to obtain the residuals. The rescaling factors correspond to the flux ratio in a FHWM aperture located at the brightest pixel in the disk GT. Axis ticks are in arcseconds.}
    \label{fig:ambig_circular}
\end{figure*}

\subsection{Sample data sets}\label{sec:sample_datasets}

To explore the disk imaging performance of the selected algorithms, we considered four empty data sets that reflect different observing conditions (see the details in Table~\ref{tab:datasets}), in which we injected five disk models with diverse morphologies (see the first row of Fig.~\ref{fig:ambig_circular}).\sj{"Added definitions and rephrased to make it intelligible} \nnewle{Specifically, we selected empty data sets from the 150 targets of the SpHere INfrared survey for Exoplanet} \citep[SHINE-F150;][]{Desidera21, Langlois21} \nnewle{taken with the Very Large Telescope (VLT) using the coronographic instrument SPHERE (Spectro-Polarimetric High-contrast Exoplanet REsearch). These data sets were obtained through} the SPHERE Data Center, with the following characteristics: low Strehl ratio with 26° rotation (ID\#1), wind-driven halo (ID\#2), unstable speckle field (ID\#3), and good Strehl ratio with 80° rotation (ID\#4). The disk morphologies were carefully chosen to represent a range of possible cases for debris and protoplanetary disks. These include two 75° inclined disks with different levels of sharpness (disks A and B), a 45° inclined disk with two concentric rings (C), one face-on disk with some azimuthal flux variation (D), and a hydrodynamical simulation computed with \texttt{Phantom} \citep{Phantom} and processed with MCFOST \citep{Pinte2006}, representing a central protostar and a low-mass companion embedded in a protoplanetary disk (E). The simulation aims to reproduce the morphology of the protoplanetary disk around MWC 758, and its exact setup is detailed in \citet{Calcino20}. \nnew{The two concentric rings in disk C were chosen to be nearly resolved. The gap between the two rings has a width of 5.3 pixels, whereas the full width at half maximum (FWHM) of our various test cubes ranges from approximately 4 to 5 pixels.}

The five synthetic disks were injected at three contrast levels ($10^{-3}$, $10^{-4}$, and $10^{-5}$), to generate a total of 60 data sets. The contrast at which the disks are injected is defined as the integrated flux measured in an \remove{FHWM}\nnew{FWHM}-sized aperture centered at the peak intensity of the disk, divided by the integrated flux \nnew{in a FWHM-sized aperture} of the point spread function. An exception was made for the synthetic disk E, where the flux is measured at the companion location. 

\subsection{Metric}
\label{sec:metrics}

To assess the quality of the disk estimations obtained with the different algorithms, we considered various metrics that are relevant to measure morphological deformations with respect to the GT image: 
\begin{itemize}
    \item the structural similarity index measure (SSIM), which captures the disparity of contrast, luminance, and structure on various windows of the two images \citep{SSIM};
    \item the Spearman's rank correlation coefficient, which assesses how well the relationship between the GT and the reconstructed images can be described using a monotonic function;
    \item the Euclidean distance, which is the square root of the sum of the squares of the pixel intensity differences between the two images;
    \item the Pearson correlation coefficient, which assesses the degree of linear correlation between the two images;
    \item the sum of absolute differences (SAD), which is calculated by summing the absolute differences between each pixel of the two images.
\end{itemize}
We selected SSIM as our preferred metric, as it is a combination of metrics covering multiple aspects of image similarity, with demonstrated performance in achieving similar assessment of structural differences as human visual perception \citep{SSIM}. However, we wish to emphasize that no single metric is better than all others in all aspects. Each metric provides different information: for example, the Spearman rank correlation coefficient better captures the global morphological difference, while the Euclidean distance is more sensitive to stellar residuals near the coronagraphic mask and/or ring-like noise artifacts in the image. Moreover, ranking disk image estimation is a multifactorial problem that often necessitates finding a satisfactory compromise between less geometrical biases and fewer noise residuals. Nevertheless, we observe that the cases where all metrics agree on the best algorithm correspond visually to disk images that are unambiguously better than others. Furthermore, despite disparities between metrics, we observe that they share common trends in their rankings.

The field of view considered for calculating the metrics also affects the results. We first calculated the metrics using a $1\arcsec$-radius aperture, excluding a 6-pixel radius inner circle located behind the coronagraphic mask.
A second set of metrics was then computed considering only pixels that are part the disk image in order to test disk reconstruction only, that      is to say, ignoring the effect of both stellar residuals near the coronagraphic mask and background artifacts on the metrics. \nnew{We summarize the results of our systematic tests in Fig.~\ref{fig:results}.} Additional figures showing which algorithm achieves the best estimation for each disk, as inferred by each metric and for these two different cases, are presented in Appendix~\ref{sec:winners} (Figs.~\ref{Fig:res_all} and \ref{Fig:res_all_disk}). 

\subsection{Results}
\label{sec:result}
We tested the performance of classical PCA, I-PCA (\texttt{GreeDS}), and \texttt{mustard} in recovering the injected disks across our 60 data sets. We optimized the parameters of I-PCA and PCA by computing a series of disk estimates with different ranks (ranging from 1 to 5), for a maximum of 10 iterations per rank in I-PCA, and selecting the estimate that performed best according to the SSIM metric. When increasing the rank in the I-PCA algorithm, we retain the last iteration of the previous rank, which means that the case \{rank=$x$, iteration=$y$\} corresponds to $(x-1)*10 + y$ iterations in total. For I-PCA, we excluded the first iteration from the selection, as it is equivalent to classical PCA.  \nnew{For \texttt{mustard}, while fine-tuning of the regularization mask from a data set to another is expected to improve our results, we instead considered a realistic scenario of not} \remove{knowing}\nnew{having access to the GT }\remove{the stellar halo}\nnew{speckle field} profile \nnew{(due to the potential presence of a disk). Hence, unlike what was shown in Fig.~\ref{fig:mask_subpot}, we did not make use of the empty data set, which is generally not accessible to the user, and rather attempted to tune the parameters of the double-Gaussian mask to minimize the distance between the data and our \texttt{mustard} model.}\remove{due to the potential presence of bright disk signals} We manually chose the regularization weights based on one data set (cube n°4, disk D, contrast $10^{-3}$), and applied these weights to all data sets. Specifically, we set the regularization percentage at initialization (see Eq.~\ref{equ:mu}) to be $p_{\rm mask} = 5\%$, $p_{\rm smooth,d} = 5\%$, and $p_{\rm smooth,s} = 5\%$.

\begin{figure*}[!t]
    \centering
    \includegraphics[width=\textwidth]{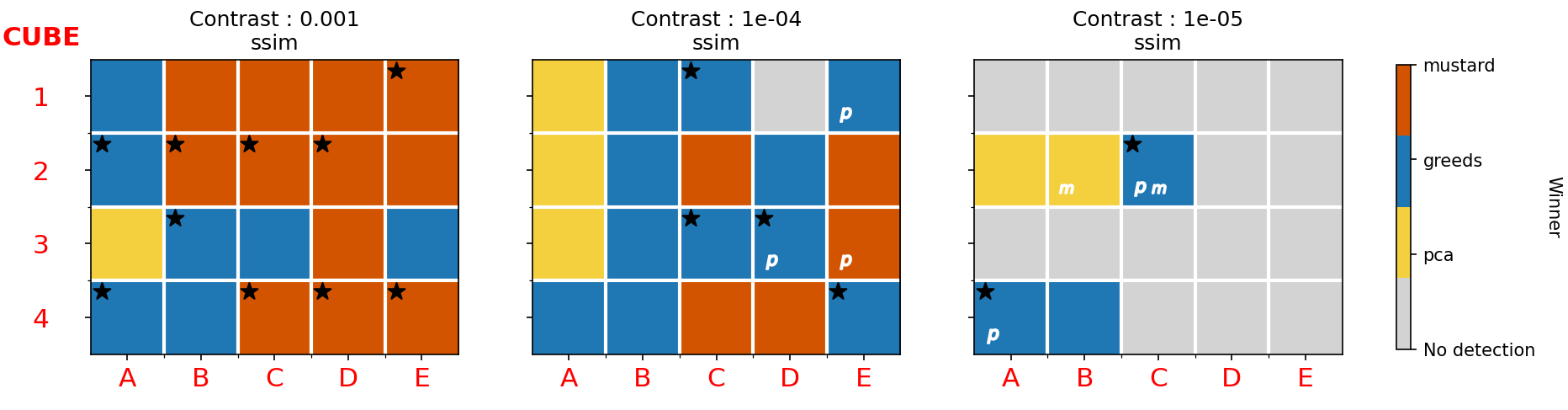}
    \caption{Results of our systematic tests to compare PCA, I-PCA (\texttt{GreeDS}), and \texttt{mustard}. Each cell represents a different synthetic data set, and its color indicates which algorithm performed best according to the SSIM metric. The five disk morphologies are labeled along the x-axis (see Fig.~\ref{fig:ambig_circular} for details), the four empty data sets used for injection along the y-axis (see Table~\ref{tab:datasets} for details), and the three contrast levels in the left ($10^{-3}$), middle ($10^{-4}$), and right ($10^{-5}$) plots. White letters on a cell indicate which algorithm(s) did not detect the disk (p stands for PCA, g for \nnew{I-PCA}, and m for \texttt{MUSTARD)}. Gray cells mean that no algorithm detected the disk. Black stars indicate that all five metrics (SSIM, Spearman, Pearson, Euclidean, and SAD) selected the same winner.}
    \label{fig:results}
\end{figure*}

In Fig.~\ref{fig:ambig_circular} we present the results of I-PCA and \texttt{mustard} for a contrast of $10^{-3}$ using data cube n°4 (a stable data set with favorable observing conditions) to highlight the main geometrical biases. To isolate these biases, we ignored flux inconsistencies by recalibrating the extracted disk images based on the integrated flux in a 1-FWHM aperture located at the maximum intensity of the GT disk, before computing the residuals. We used a color scale where blue indicates missing signal and red indicates excess. We observe that missing signal in I-PCA estimates is symptomatic of flux invariant to rotation. Disks A and B only show missing flux close to the coronagraphic mask, while disks C, D, and E show more flux invariant to rotation, leading to more significant deformations. The \texttt{mustard} IP approach can significantly reduce geometric biases related to flux invariant to rotation. However, it shows more noise residuals, especially in the region close to the coronagraphic mask and at the edge of the well-corrected region by the SPHERE deformable mirror (also known as the control ring), which may be due to the mask $M_{\mathrm{reg}}$ and/or the balance of regularization not being appropriate to the morphology of the speckle field. The sub-optimality of $M_{\mathrm{reg}}$ could have been corrected by estimating the mask from the empty data cube, but this would be unfair to the other algorithms. In the future, adapting the \texttt{mustard} ADI algorithm using a reference star as a data-driven prior is a potential avenue of improvement, as discussed in Sect.~\ref{sec:conc}.

\remove{We summarize the results of our systematic tests in Fig.~\ref{fig:results}.} The disk reconstructions, along with the names of the best algorithms, as inferred by each metric, are presented in Appendix~\ref{sec:simus}. 
Regardless of the chosen metrics (see Appendix~\ref{sec:winners}), we observe that \texttt{mustard} can provide good results in the most favorable cases, namely bright disks in data sets with a close to static speckle pattern. \texttt{Mustard} performs particularly well for disks that contain more flux that is invariant to rotation (i.e., disks C, D, and E, the more face-on disks from our samples). In contrast, I-PCA is more robust in more challenging cases, performs better than simple PCA in most cases, and outperforms \texttt{mustard} in most cases where either the speckle field is unstable or the contrast is deeper than $10^{-4}$. We also note that the sharp-edged disk (A) is systematically better estimated by I-PCA than \texttt{mustard}. This is likely due to the minimal amount of rotation-invariant flux in this disk, which means that its overall morphology is not much distorted with I-PCA, with the resulting image containing less background noise and stellar residuals near the \remove{coronographic}\nnew{coronagraphic} mask than the one obtained with \texttt{mustard}.

\begin{figure}[t]
    \centering
    \includegraphics[width=\linewidth]{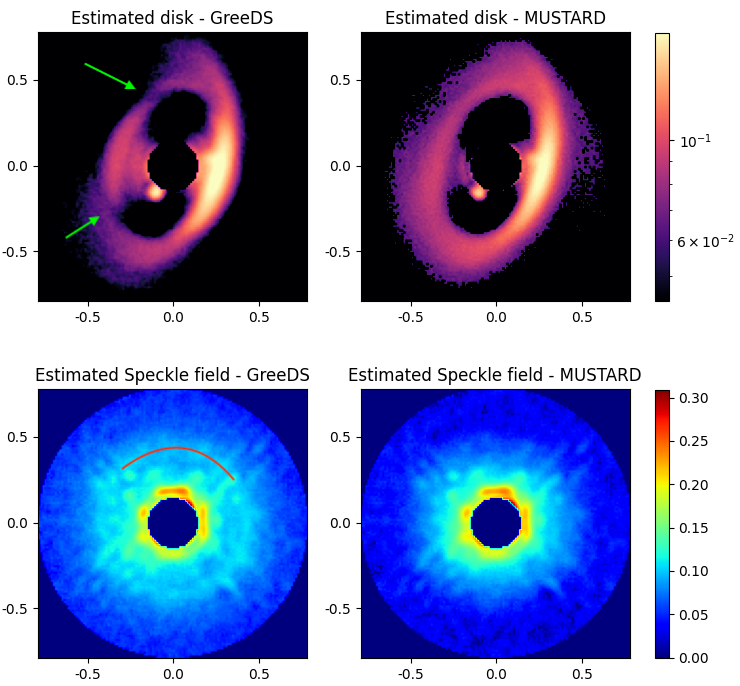}
    \caption{Images of a VLT/SPHERE/IRDIS $\nnew{K1}$ data set on PDS 70 obtained using I-PCA (left) and \texttt{mustard} (right). The top row displays the estimated circumstellar signal, while the bottom row shows the mean estimated \remove{stellar PSF}\nnew{speckle field}. The color scale is arbitrary (yet consistent between all plots in relative terms) and displayed in a logarithmic scale. The green arrows indicate the location of shadows in the disk, which are the consequences of the lack of circular component in the disk, and are consistent with the location of a bright arc (highlighted in red) in the speckle field estimation.}
    \label{fig:PDS}
\end{figure}
\begin{figure}[!t]
    \centering
    \includegraphics[width=\linewidth]{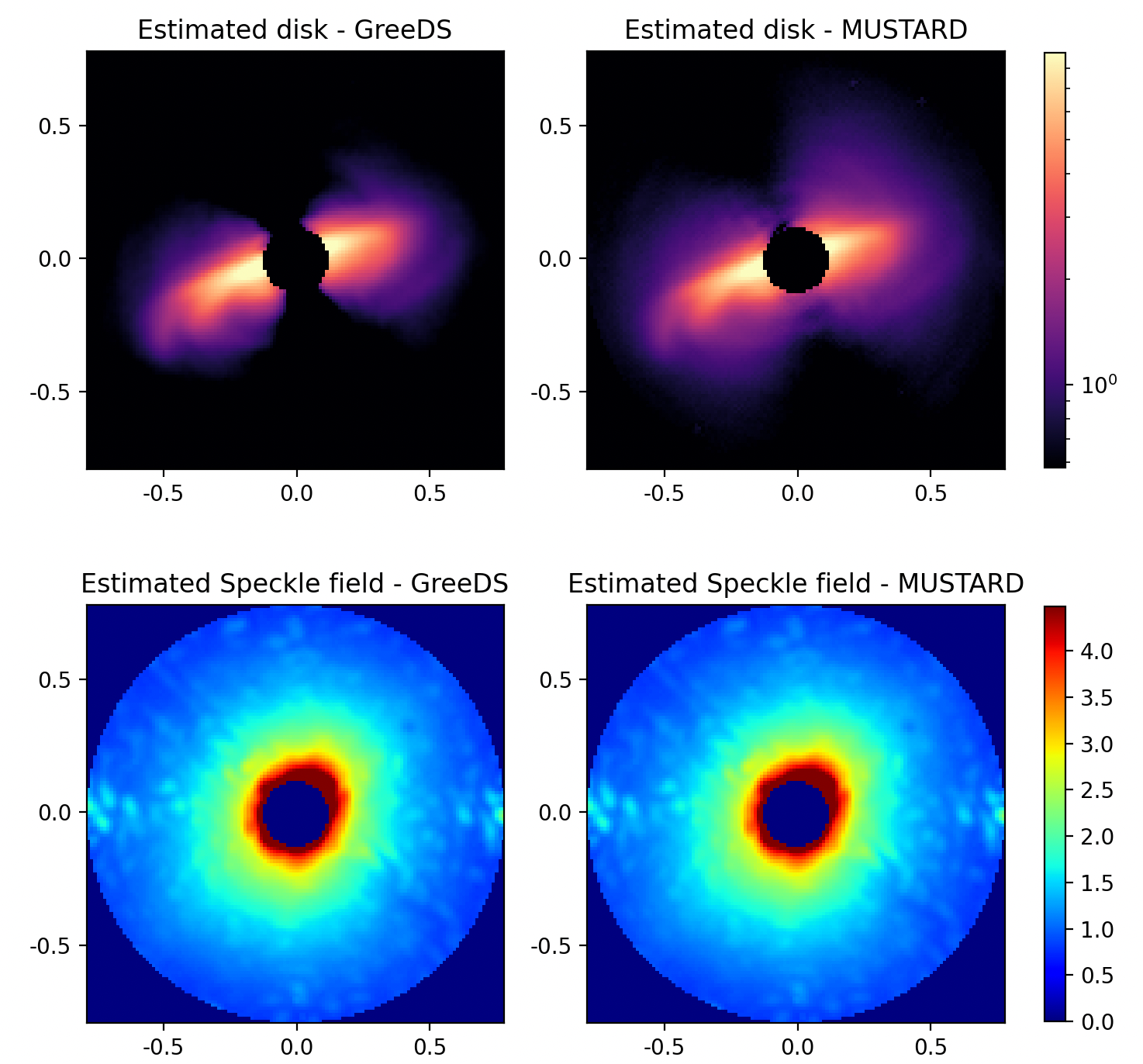}
    \caption{Same as Fig.~\ref{fig:PDS} but for a VLT/SPHERE/IRDIS $\nnew{H2}$ data set obtained on RY Lupi (program ID 097.C-0865, 2016).}
    \label{fig:rylup}
\end{figure}

\subsection{Tests on real data}
\label{sec:realdata}

Here, we present a comparative analysis of disk imaging for two ADI data sets already processed with \texttt{MAYO} and/or \texttt{REXPACO} in the literature. The first data set was obtained on PDS~70 (ESO program ID 1100.C-0481) \nnew{in the K1 band} with the infrared dual-band imager and spectrograph (IRDIS) \LEt{ Consider defining. ***}camera of the Spectro-Polarimetric High-contrast Exoplanet REsearch coronographic system of the Very Large Telescope (VLT/SPHERE) in March 2019, under excellent conditions (seeing: $0\farcs39$, wind speed: 4.45 m/s, $\Delta_{\theta}=56\fdg5$). This data set had previously been processed using PCA (5 PCs) in \cite{periodPlanetc}. The same data set was processed with the \texttt{MAYO} and \texttt{REXPACO} algorithms, respectively in \citet{mayo} and \citet{REXPACO}. The second ADI data set was obtained with the same instrument and camera on RY~Lupi (HIP~78094) in March 2016 (ESO program ID 097.C-0865) \nnew{in the H2 band} under excellent conditions (seeing: $0\farcs25$, Strehl ratio 78\%, $\Delta_{\theta}=71\degr$). The data set was reduced using cADI and no-ADI (averaging after de-rotation without subtraction) in \cite{Langlois18}, and with \texttt{REXPACO} in \citet{REXPACO}. The reader is referred to the original papers for the \texttt{MAYO} and \texttt{REXPACO} results, which we do not reproduce here for the sake of conciseness. In all cases, we observe that the \texttt{MAYO} and \texttt{REXPACO} algorithms, as well as \texttt{mustard} and I-PCA (see below), perform better than the original cADI or PCA reductions at restoring extended signals. 

We processed the PDS~70 data set using \texttt{mustard} and I-PCA, and we compare in Fig.~\ref{fig:PDS} the disk estimation and the averaged estimated speckle field obtained with these two algorithms. In the speckle field estimation of I-PCA, we observe a suspicious arc at a radius of $\sim 0\farcs4$, which matches the location of shadows in the disk estimate (red arc and green arrows in Fig.~\ref{fig:PDS}, respectively). This indicates that flux invariant to rotation is likely missing from the disk estimate, which could bias the interpretation of the final image, and potentially enhance an arm-like structure observed to the northwest. We observe the same signpost of missing flux invariant to rotation in the \texttt{MAYO} images \citep[Fig.~9 in][]{mayo} and in the \texttt{REXPACO} images \citep[Fig.~11 and 12 in][]{REXPACO}. Although \texttt{MAYO} performed deconvolution of the signal and separated point-like sources from the disk signal, we observe that the shearlet transform appears to have fitted the deformation inherited from I-PCA instead of correcting it. The \texttt{MAYO} image indeed suggests two point-like features (PLFs), among which PLF 2, at the east of the coronagraphic mask \citep[Fig.~9 in][]{mayo}, likely corresponds to a filtered spiral-like signal (or to bright stellar residuals near the coronagraphic mask).
Based on our \texttt{mustard} images, we propose an alternative interpretation of the disk: the overall shape is closer to a regular ellipse, without shadows, and the arm-like structure to the northwest is real, but less prominent. This is inferred solely from the shape of the average speckle field. In \citet{Sandrine}, we confirmed that the observed structure exists at the northwest and deviates from what we would expect for a double-ring disk.

The disk and average speckle field estimates obtained using I-PCA and \texttt{mustard} for RY Lupi are presented in Fig.~\ref{fig:rylup}. The primary ring of the disk is correctly imaged by all algorithms, but the spiral-like features \citep{Langlois18} are better restored with I-PCA, \texttt{mustard} and \texttt{REXPACO} \citep[Fig.~11 and 12 in][]{REXPACO}. The cADI \citep[Fig.~1 in][]{Langlois18} is too aggressive toward these fine structures, while the no-ADI reduction hides them in the stellar halo. Compared to I-PCA, the interpretation of the disk by \texttt{mustard} does not provide much additional information because, unlike PDS~70, RY~Lupi does not contain much bright signal that is ambiguous with respect to the $71\degr$ field rotation.

\section{Discussion}
\label{sec:discussion}

We conducted a comprehensive evaluation of state-of-the-art algorithms for processing ADI data sets containing disks, aiming to determine their respective scope of validity. Our analysis involved systematic testing of PCA, I-PCA (\texttt{GreeDS}), and \texttt{mustard} on a diverse set of 60 synthetic data sets, encompassing various observing conditions and disk morphologies. Additionally, we compared the performance of these algorithms on a selection of real data sets. This comparative study enables us to analyze the strengths and limitations of the \texttt{mustard} IP approach with respect to PCA-based methods, focusing on three key aspects: (i) theoretical faithfulness, (ii) results, and (iii) automation/practicality (see Sect.~\ref{sec:ip-vs-pca}). Subsequently, in Sect.~\ref{sec:comparison}, we explore the distinctions between the state-of-the-art IP algorithms and assess their respective use cases, identifying scenarios where one algorithm may outperform the others.

\subsection{Static speckle model versus PCA-based estimation} \label{sec:ip-vs-pca}

In terms of faithfulness, no algorithm can provide a deformation-free disk due to the flux invariant to rotation and the absence of a perfect model to describe the quasi-static behavior of the speckle field. As seen in Fig.~\ref{fig:ambig_circular}, the limitation associated with flux invariant to rotation can significantly impact disk morphologies, even in inclined disks. \texttt{Mustard} is the only algorithm that will not systematically exclude rotation-invariant flux from the disk estimate, but the fate of this component will still depend on the chosen prior ($M_{\rm reg}$). Apart from this issue, the IP approach employed by \texttt{mustard} and \texttt{REXPACO} offers improved theoretical faithfulness compared to PCA-based methods. This is achieved by leveraging a concrete and constrained model of the ADI cube. However, this approach relies on the assumption that the speckle field remains static, meaning that there are minimal relative variations in the \remove{stellar PSF}\nnew{speckle field} with respect to disk signal. In other cases, PCA-based methods can in practice be a good approximation to model the speckle field, and hence a relatively robust disk image, despite not providing any theoretical faithfulness. 

In terms of results, we observe that I-PCA performs significantly better than PCA for restoring extended signals regardless of cube quality, contrast, and disk morphology \citep[as also noted in][]{StapperGinski22}. With an increasing number of iterations, more disk signal is recovered, but more noise residuals are also propagated and amplified, leading to strong background artifacts when using a large number of iterations. One example is the estimation of Disk A in Cube 2 with a contrast of $10^{-5}$ (as shown in Fig.~\ref{fig:2A}), which was processed using a rank of three and one iteration, totaling $2\times10 + 1 = 21$ iterations (see Sect.~\ref{sec:result}). We note \LEt{ or "It should be noted" or similar -- A\&A encourages authors to avoid directly addressing the reader in the main text, e.g., by using the imperative or posing direct questions. ***}that our iterative optimization method and I-PCA implementation may not be optimal in all cases: for example, high-contrast disk estimates may be improved by starting to iterate at a higher rank. A better optimization of I-PCA, including additional strategies such as thresholding, masking, or using temporal-PCA, could still improve the results %and should be explored in a specific study such as 
\citep{StapperGinski22, Long23}. In general, we observe that \texttt{mustard} is the best algorithm in the most favorable cases (bright disk, stable data sets), achieving both minimal reconstruction errors and optimal flux estimation, especially for disks that contain more flux invariant to rotation as \texttt{mustard} \nnewle{partially assigns the ambiguous flux to}\removele{does not exclude flux from} the disk estimate, according to a prior\sj{ Meaning changed: The flux invariant to rotation is not excluded; this type of flux can contribute to the estimated disk. We rephrased it in hopes of making it clearer}\LEt{ Verify that your intended meaning has not been changed. ***}. However, \texttt{mustard} does not perform well at high contrasts, and is more prone to contain circularly invariant artifacts, especially near the coronagraphic mask region, which explains why it is less often selected as the best reconstruction algorithm by metrics calculated on the whole field of view than calculated on the pixels of the disk only (see Appendix~\ref{sec:winners}). These artifacts are caused by the inappropriate assumption about the static behavior of the speckle field, and the imperfect regularization mask. \nnew{Indeed, in addition to the difficulty in estimating the parameters for shaping the mask, we can observe in Fig~\ref{fig:mask_subpot} that the double Gaussian assumption appears inappropriate for capturing the innermost region, as well as the bright speckle ring at a separation of $\sim 0\farcs7$ (i.e., the edge of the region corrected by the deformable mirror).} \remove{to sort the stellar halo from the circularly invariant component araising from the disk;} Consequently, for high relative variations in the \remove{stellar PSF}\nnew{speckle field} with respect to disk signal, the disk gets drowned in the stellar residuals.

In terms of practicality, automating the process of setting hyperparameters is a known limitation of IP approaches in general.
In the case of I-PCA, finding the correct number of iterations and rank is crucial as the algorithm does not guarantee convergence for an increasing number of iterations, making it unclear when to stop and which image to consider as the final result in a real-world scenario. Despite this, compared to IP approaches, I-PCA is more intuitive, faster, and easier to parametrize, and the evolution of frames generated through the iteration can be used to facilitate the process. \nnew{Additionally, another factor to account for in terms of algorithm practicality is computation time, which heavily relies on the implementation approach. The \texttt{mustard}, \texttt{MAYO}, and \texttt{GreeDS} algorithms (assuming ten iterations per rank for ten ranks) are implemented using the PyTorch package in Python. For a data set comprising 100 frames with dimensions of $256\times256$, the computational duration for the three algorithm implementations is roughly 20 minutes, 1 day, and a few seconds, respectively (2,8 \LEt{ should this be 2.8?. ***}GHz Intel Core i7 CPU; 16Go RAM). Alternative implementations of I-PCA utilizing NumPy might require a few minutes to process.}

\subsection{Comparison of IP approaches} \label{sec:comparison}

It is challenging to make a fair comparison of empirical results between different IP approach algorithms, such as \texttt{REXPACO}, \texttt{MAYO}, and \texttt{mustard}, due to inconsistencies in the test methodology used in the respective publications. The quality of the data cubes in which the synthetic disks are injected, including the Strehl ratio, variability of the speckle field, amplitude of rotation, or presence of wind-driven halo, can significantly influence the results. Additionally, the definition of contrast for disks may not be equivalent in all publications. In this paper, we find that disks with contrasts $\sim 10^{-5}$ were hardly detectable, whereas in \cite{REXPACO}, cADI was able to recover disks with similar contrasts reasonably well, even though cADI is not known to perform better than PCA \citep{DC1}. Similarly, \cite{mayo} claim that it is possible to detect disks with contrasts as deep as $10^{-9}$, which is outside the range of contrast typically tested for algorithms thus far, for both disks and planets. \nnew{Moreover, evaluating contrast at a single peak point, as performed in this paper, might not accurately depict the contrast of extended sources, especially for disks that exhibit a high peak that gradually diminishes in intensity as one moves away from the peak. More detections would have been possible if the intensity was uniform across all pixels of the disk. As a result, the ambiguity of this contrast definition persists.} Therefore, only a standardized test pipeline such as the Exoplanet Imaging Data Challenge \citep{DC1,DC2} would provide a fair empirical test to compare different IP approaches. One can still highlight specific features in the results obtained by these various algorithms, and make some recommendations regarding their usage. %Peut-être ajouter une recommandation sur l'utilisation de MUSTARD ici?

A specific feature of \texttt{MAYO} is that it includes an interesting interpretation of the deconvolved disk and planet. However, its model lacks identifiability (see Sect~\ref{sec:ip-comp}). When using \texttt{MAYO} to retrieve disks from ADI sequences, one should systematically check the \nnew{I-PCA} first estimate \nnew{provided by \texttt{GreeDS}} to locate potential artifacts and ambiguities that could be overfitted. Due to the lack of statistical validity, we observe that the algorithm tends to fit deformations rather than correcting them (see Sect.~\ref{sec:realdata}). Consequently, in terms of faithfulness, \texttt{MAYO} does not provide a significant improvement over \texttt{GreeDS}. 

Unlike other algorithms, \texttt{mustard} specifically tackles the issue of flux invariant to rotation. It provides a less aggressive interpretation of the disk that keeps more flux invariant to the rotation according to a prior, using a mask that has to be adapted by the user. However, it is limited by our imprecise knowledge of the speckle field, and the arbitrary choice of the mask $M_{\rm reg}$ and regularization weights $\mu$ may significantly alter the results. Nevertheless, \texttt{mustard} is particularly useful for disks that have a significant amount of flux invariant during rotation, and are brighter than the variations in the speckle field.

In terms of automation, each IP approach proposes an automated or semiautomated method for setting the hyperparameters. \texttt{REXPACO} and \texttt{MAYO} propose automated methods that search for optimal hyperparameters by minimizing the data-attachment term using regression algorithms. However, these techniques can be time-consuming and assume that the data-attachment term provides an ideal description of the data, which is not always the case (see Sect.~\ref{sec:ip-comp}). Consequently, achieving the global minimum of the data-attachment term might not necessarily result in the optimal disk estimate. On the other hand, \texttt{mustard} suggests defining regularization as a percentage of the data-attachment term. While this approach lacks statistical guarantees on the results, it can provide users with more intuitive parameter setting capabilities. %Finally, each IP approach, propose an automated or semi-automated method to set the hyperparameters.
 
% Factors such as the variability of the speckle field, imperfection of the PCs, wind-driven halo, and other phenomena are beyond the scope of validity of the model.

\section{Conclusion}
\label{sec:conc}

We have introduced \texttt{mustard}, an IP approach designed to address the recovery of flux invariant to rotation in circumstellar disk images, which is an inherent limitation of the ADI observing strategy. \removele{The algorithm that specifically addresses this limitation is \texttt{mustard}, which provides}\nnewle{No algorithm had specifically addressed this limitation before \texttt{mustard}, which hence provides} an alternative interpretation of disk imaging data sets, with a better preservation of the flux invariant to rotation. We have also provided a comprehensive comparison of state-of-the-art algorithms to retrieve extended disk signals from ADI sequences and discuss their advantages and limitations. \texttt{Mustard} is suitable for stable data sets that fall within its scope of validity. However, it exhibits limited robustness when confronted with large relative variations in the\remove{stellar PSF} \nnew{speckle field} compared to the disk signal. Iterative PCA proves to be more resilient in handling high-contrast scenarios, unstable speckle fields, and mild wind-driven halos. Our analysis suggests that, in most cases, I-PCA outperforms \texttt{mustard} and should be the preferred choice for recovering disks from ADI sequences. This recommendation is based on I-PCA's combined performance and practicality, even though it does not completely address the challenges of flux invariant to rotation, nor provide statistical guarantees. Considering these factors, I-PCA serves as a solid initial approach to take before considering more sophisticated methods.
Nevertheless, \texttt{mustard}, \texttt{REXPACO}, and \texttt{MAYO} can be utilized when their scope of validity and purpose align with the specific application scenario. For instance, \texttt{mustard} is particularly useful for recovering rotation invariant flux, while \texttt{REXPACO} offers enhanced guarantees and \texttt{MAYO} excels in deconvolution tasks. The choice of method depends on the specific requirements and goals of the application at hand.

An enhancement to a model-based approach such as \texttt{mustard} could be achieved by incorporating  a mathematical model that describes the temporal evolution and morphology of the speckle field and accounts for known nuisance phenomena, such as the wind-driven halo. These limitations currently impact the quality of the model, resulting in challenges when recovering disks at higher contrasts. Therefore, there is a need to develop a more realistic and constrained model that effectively correlates with the observed data, addressing these limitations and improving the overall performance of \texttt{mustard}. \nnew{Ambiguities related to flux invariance during rotation are a limitation specific to ADI observing strategies. Approaches that incorporate reference stars (star hopping/RDI\LEt{ slash. ***}) do not face this particular ambiguity. Some of the most notable efficiency gains compared to the PCA/median \LEt{ slash. ***}subtraction technique were obtained by leveraging RDI \LEt{ Please define. ***}and imposing non-negativity constraints on the components employed to represent the estimated speckle field, through non-negative matrix factorization of the reference images, and either a concurrent search for an optimal scaling factor of the model speckle field or data imputation \citep[][]{Ren18, Ren20}. %Additionally, the post-processing NMF algorithm enables the optimal scaling of the speckle field model derived from reference stars. This optimal scaling, applied prior to subtracting the model from the data, effectively mitigates the issue of over-subtraction.
}

\nnew{Moreover, combining diverse observing strategies might assist in mitigating the distinct weaknesses of each approach, as exemplified by \citet{Lawson22}, who merged RDI with PDI\LEt{ Please write out. ***}, or by \citet{Flasseur22}, who merged ADI with SDI\LEt{ Please write out. ***}.} \remove{Hence, a promising approach to simultaneously solve the poor handling of circularly invariant signals and the lack of substantiated models for the stellar halo in ADI sequences, is to leverage a library of reference stars.}\nnew{Therefore, to address the inherent ambiguities of ADI, we propose exploring the combination of ADI with RDI in a future work.} Reference stars provide reliable data-driven priors that can be used to solve the problem of the flux invariant to rotation. The combination of ADI with reference stars could, theoretically, offer the advantages of both methods: improving the correlation between the estimated speckle field through ADI and solving ambiguities that limit angular diversity through the use of reference stars. This can improve both I-PCA methods, by concatenating reference frames in the cube to compute the PCs, and IP approaches, by building the optimal prior through regularization based on the reference data. This will be the subject of future work (Juillard et al., in prep).

%% ---------------------------- REFS and ABSTARCT ------------------------------------------- %%

\begin{acknowledgements}
This project has received funding from the European Research Council (ERC) under the European Union's Horizon 2020 research and innovation program (grant agreement No 819155), and from the Belgian Fonds de la Recherche Scientifique -- FNRS. This work has made use of the SPHERE Data Centre, jointly operated by OSUG/IPAG (Grenoble), PYTHEAS/LAM/CESAM (Marseille), CA/Lagrange  (Nice), Observatoire de Paris/LESIA(Paris), and Observatoire de Lyon \citep{Galicher, Delorme}. We are very grateful to Benoit Pairet and Laurent Jacques for their feedback on this work and useful discussions, and to Julien Milli, Philippe Delorme, and Mariam Sabalbal for their help in selecting, providing, and pre-processing the empty SPHERE data cubes used for the tests. 

\end{acknowledgements}

\bibliographystyle{aa} % style aa.bst
\bibliography{references}

\begin{appendix}

\section{Comparison of results for different metrics }
\label{sec:winners}

\begin{figure}[!t]
    \centering
    \includegraphics[width=\linewidth]{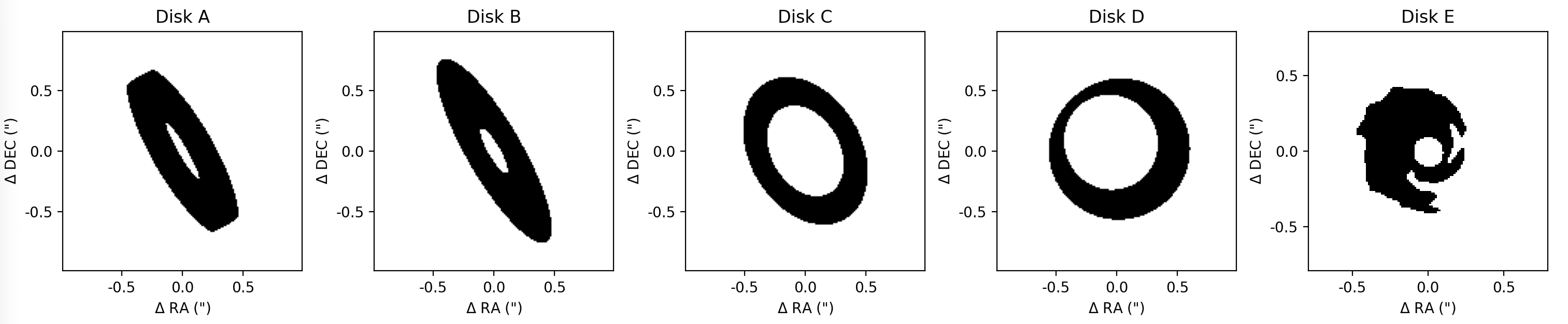}
    \caption{Mask used to compute metrics and considering only pixels of the disk.}
    \label{fig:mask85}
\end{figure}

\begin{figure*}[p]
    \centering
    \includegraphics[width=0.9\linewidth]{Images/table_ssim.png}
    \includegraphics[width=0.9\linewidth]{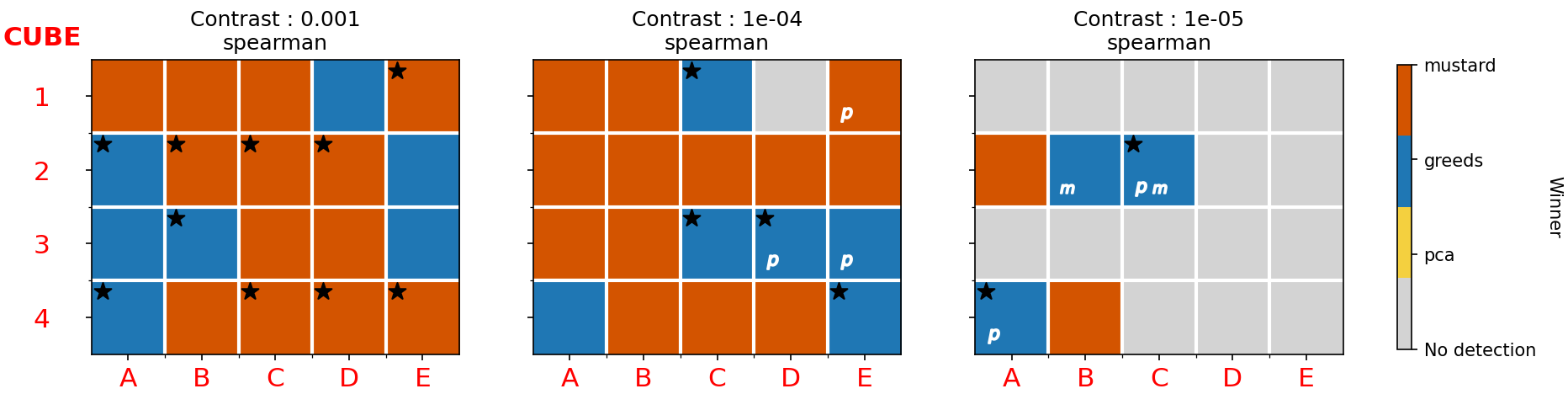}
    \includegraphics[width=0.9\linewidth]{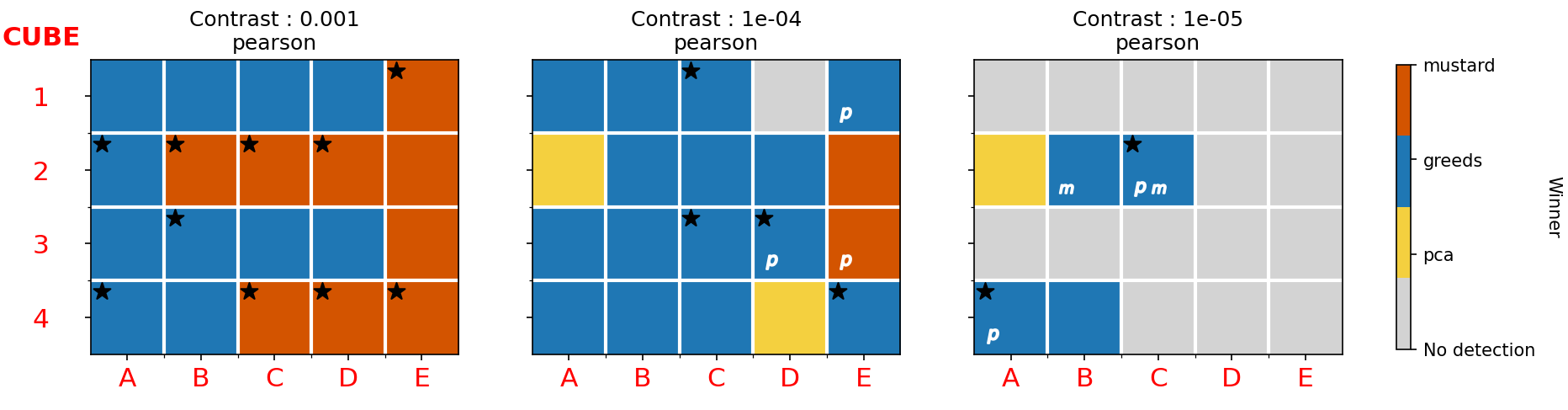}
    \includegraphics[width=0.9\linewidth]{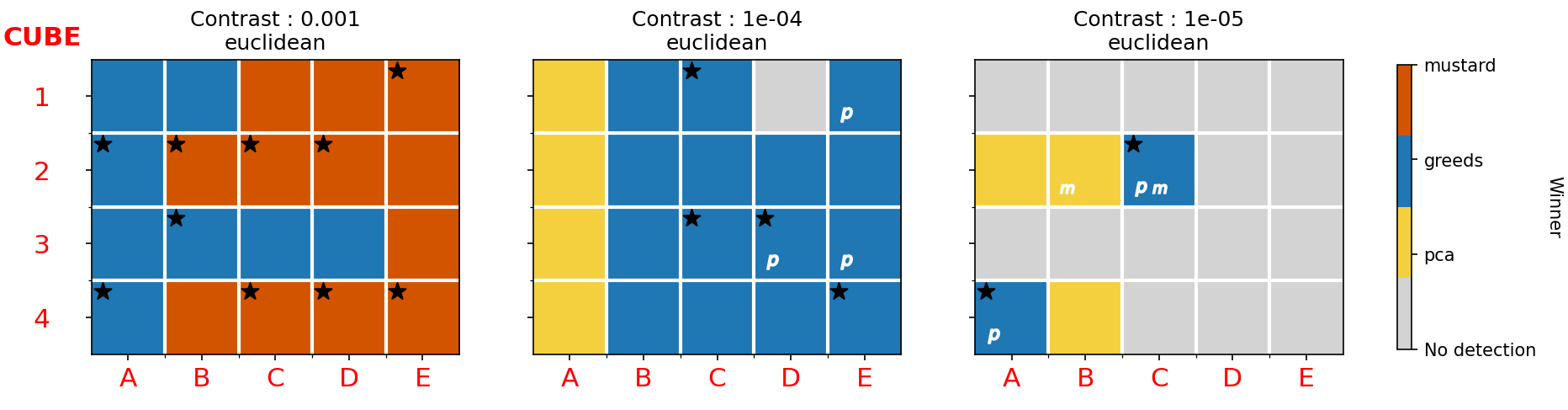}
    \includegraphics[width=0.9\linewidth]{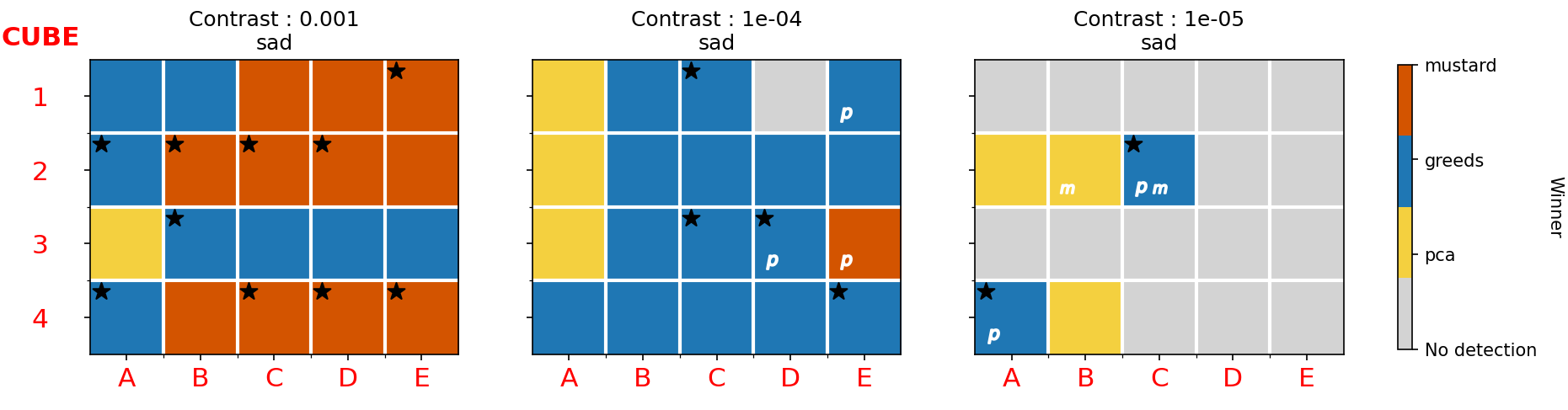}
    \caption{Same as Fig.~\ref{fig:results} but for the five different metrics defined in Sect.~\ref{sec:metrics} (from top to bottom): SSIM, Spearman's rank, Pearson correlation, Euclidean distance, and SAD distance.}
    \label{Fig:res_all}
\end{figure*}

\begin{figure*}[p]
    \centering
    \includegraphics[width=0.9\linewidth]{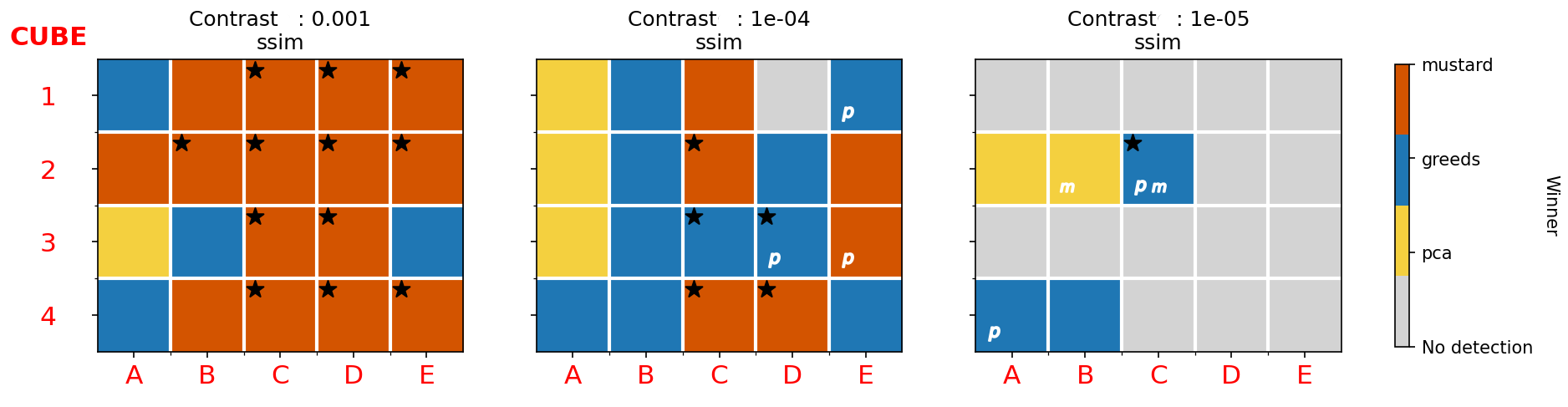}
    \includegraphics[width=0.9\linewidth]{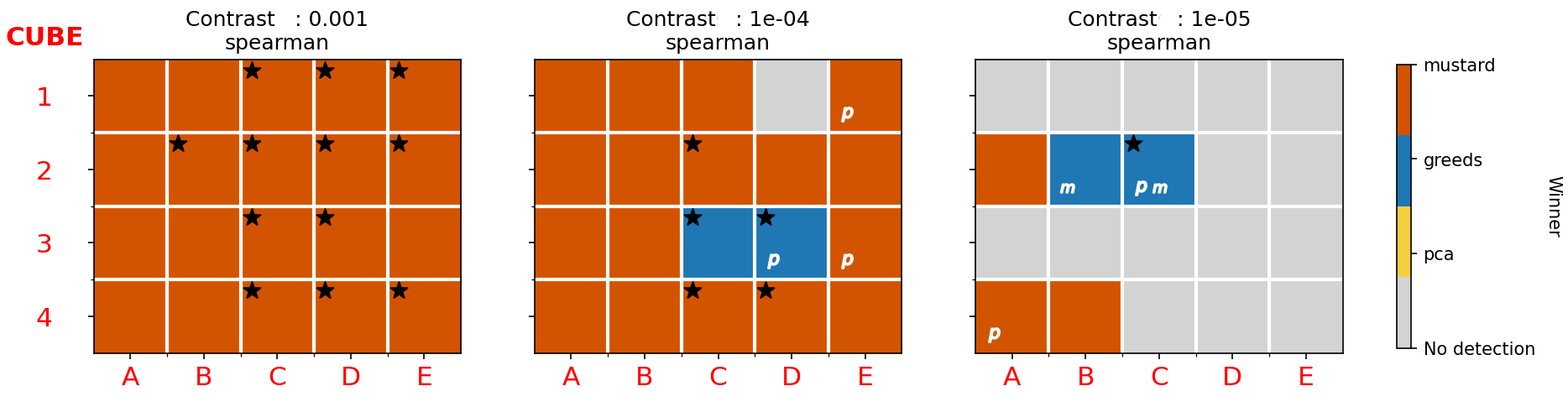}
    \includegraphics[width=0.9\linewidth]{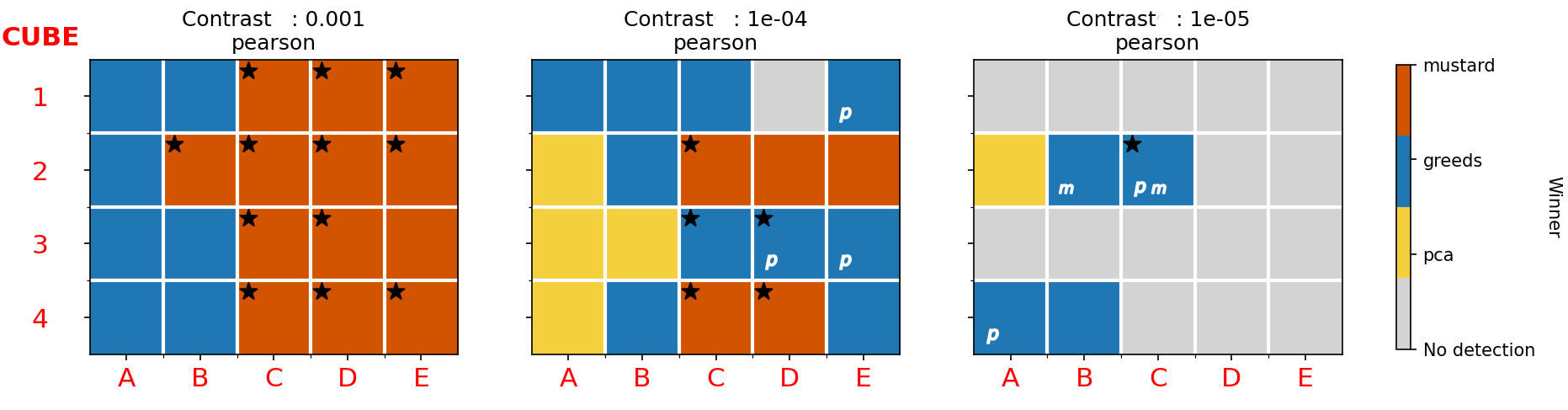}
    \includegraphics[width=0.9\linewidth]{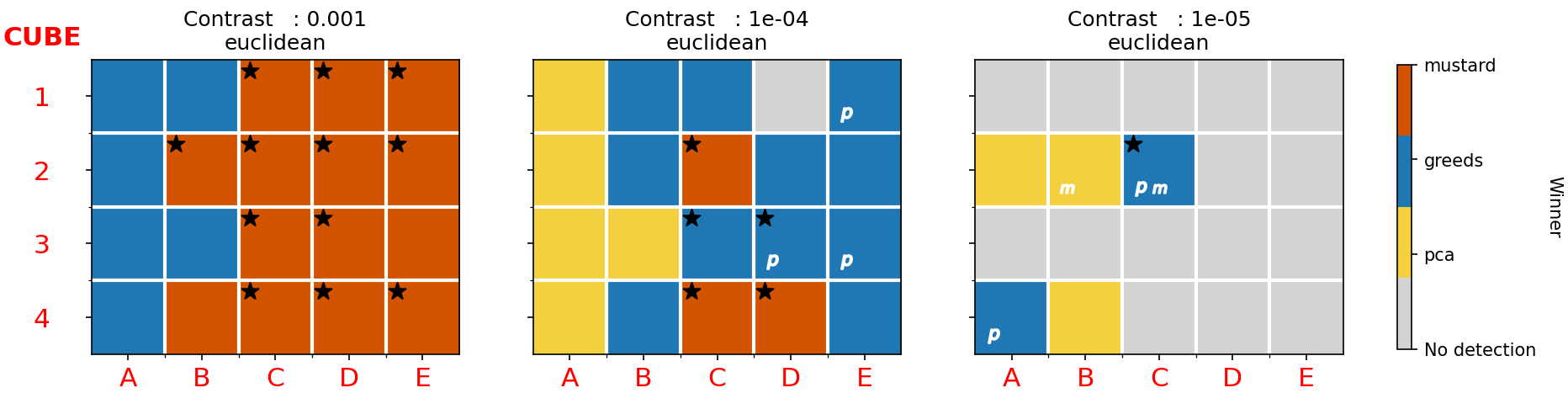}
    \includegraphics[width=0.9\linewidth]{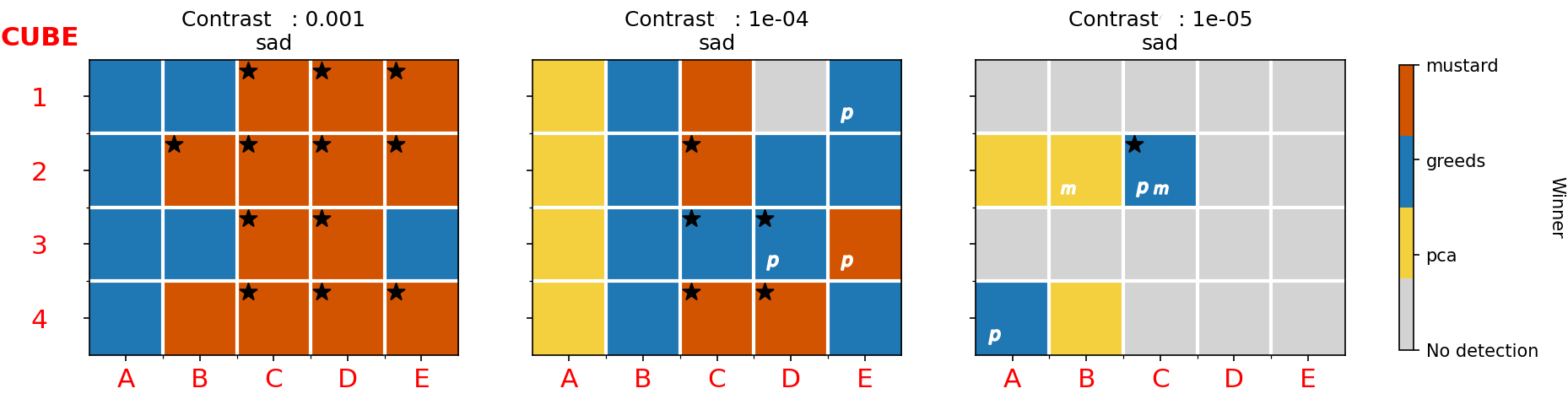}
    \caption{Same as Fig.~\ref{Fig:res_all}, but only considering the pixels that are part of the disk image.}
    \label{Fig:res_all_disk}
\end{figure*}

This Appendix presents the results of the performance comparison for the three considered algorithms (PCA, I-PCA (\texttt{GreeDS}), and \texttt{mustard}) in terms of the inferred disk image, for each test data set described in Sect.~\ref{sec:sample_datasets}, using the five different metrics described in Sect.~\ref{sec:metrics}: the SSIM \citep{SSIM}, the Spearman's rank correlation coefficient, the Pearson correlation coefficient, the Euclidean distance, and the SAD.
Two sets of results are presented. The first set of results in Fig.~\ref{Fig:res_all} shows the metrics computed considering a $1\arcsec$-radius field of view, while the second set of results in Fig.~\ref{Fig:res_all_disk} shows metrics calculated on disk pixels only, defined considering the 85 percentile of the GT disk's intensity (Fig~\ref{fig:mask85}).

For each subplot within these two figures, the cells represent different test data sets. The color of the cell indicates which algorithm performed the best according to the respective metrics. We simulated five different disk morphologies \textbf{(x-axis, A→E)}, injected them in four different empty data sets of stars observed with the IRDIS camera of the VLT/SPHERE instrument under various conditions \textbf{(y-axis, 1→4)}, and tested our three algorithms for injections at three different contrast levels: $10^{-3}$ \textbf{(left)}, $10^{-4}$ \textbf{(middle)}, and $10^{-5}$ \textbf{(right)}. The five different disk morphologies are illustrated in Fig.~\ref{fig:ambig_circular}, while details on the four empty data sets can be found in Table~\ref{tab:datasets}. Gray cells mean that no algorithm detected the disk. White letters on a cell indicate which algorithm(s) did not detect the disk (p$\longrightarrow$PCA, g$\longrightarrow$\remove{\texttt{GreeDS}}\nnew{I-PCA}, m$\longrightarrow$\texttt{MUSTARD)}. Black stars indicate that all five metrics (Spearman, SSIM, Euclidean, Spearman, Pearson, and SAD) agree on the algorithm which achieved the best disk estimation.

From these two figures, it can be observed that the various metrics yield relatively similar results. \texttt{Mustard} is more frequently chosen as the best reconstruction by all metrics for low contrasts ($10^{-3}$) and for the more face-on disk (C, D, E). However, when considering the entire image, I-PCA is preferred in most cases, with an average selection rate of 53\% across all metrics (considering detection only). In comparison, \texttt{mustard} is selected on average 37\% of the time, while PCA is selected only 10\% of the time. We note that these rates exclude the non-detections, which represent 26\% of our 60 simulations. When considering only the disk pixels and using the Spearman's rank correlation coefficient, which are better suited for assessing the preservation of global morphology and are less sensitive to background noise and image residuals, the results generally favor \texttt{mustard}.

 \section{Results of disk estimations}
\label{sec:simus}

This Appendix presents the disk estimations obtained with PCA, I-PCA (\texttt{GreeDS}) and \texttt{mustard} for all 60 test data sets considered in this work (Figs.~\ref{fig:4E}--\ref{fig:1A}). Each page-sized figure presents the results for a specific synthetic disk in a given data sets. % Details on the different empty data sets (1→4) can be found in Table~\ref{tab:datasets}, while the injected disks (A→E) are shown in the left-most column of each Figure. Each figure contains three pairs of rows, each corresponding to a different injected contrast level: $10^{-3}$ (top), $10^{-4}$ (middle), and $10^{-5}$ (bottom). The top row of each pair represents the estimation, while the bottom row represents the residuals. Each column displays the following in order: ground truth (left), PCA (middle-left), I-PCA (middle-right), and \texttt{mustard} (right). The best method, as determined by each metric, is indicated in the bottom-left corner of each pair of rows.

The color bar bounds in the residual plots is set to $\pm$ maximum of intensity of the GT, centered at 0. For the disk image estimation plots, the color bar is adjusted for each data set to have a maximum value equal to the 99th percentile of the image. The minimal value is set to 0. Additionally, the optimal PCA and I-PCA parameters (rank and number of iterations) are written at the top left of the corresponding images.

Page-sized figures for each data set are ordered from the easiest data cube (ID\#4, stable) to the hardest (ID\#1, low Strehl ratio), and from the most face-on disk (disk E) to the sharpest one (disk A), in order to illustrate the most significant improvements brought by the \texttt{mustard} model in the first pages of this annex.

\def\fdisk{E}
\def\fcube{4}
\foreach \cube in {4, 3, 2, 1}
    \foreach \disk in {E, D, C, B, A}
    {
    \ifx \cube\fcube
        \ifx \disk\fdisk
        \begin{figure*}[!ht]
            \centering
            \includegraphics[width=0.95\linewidth]{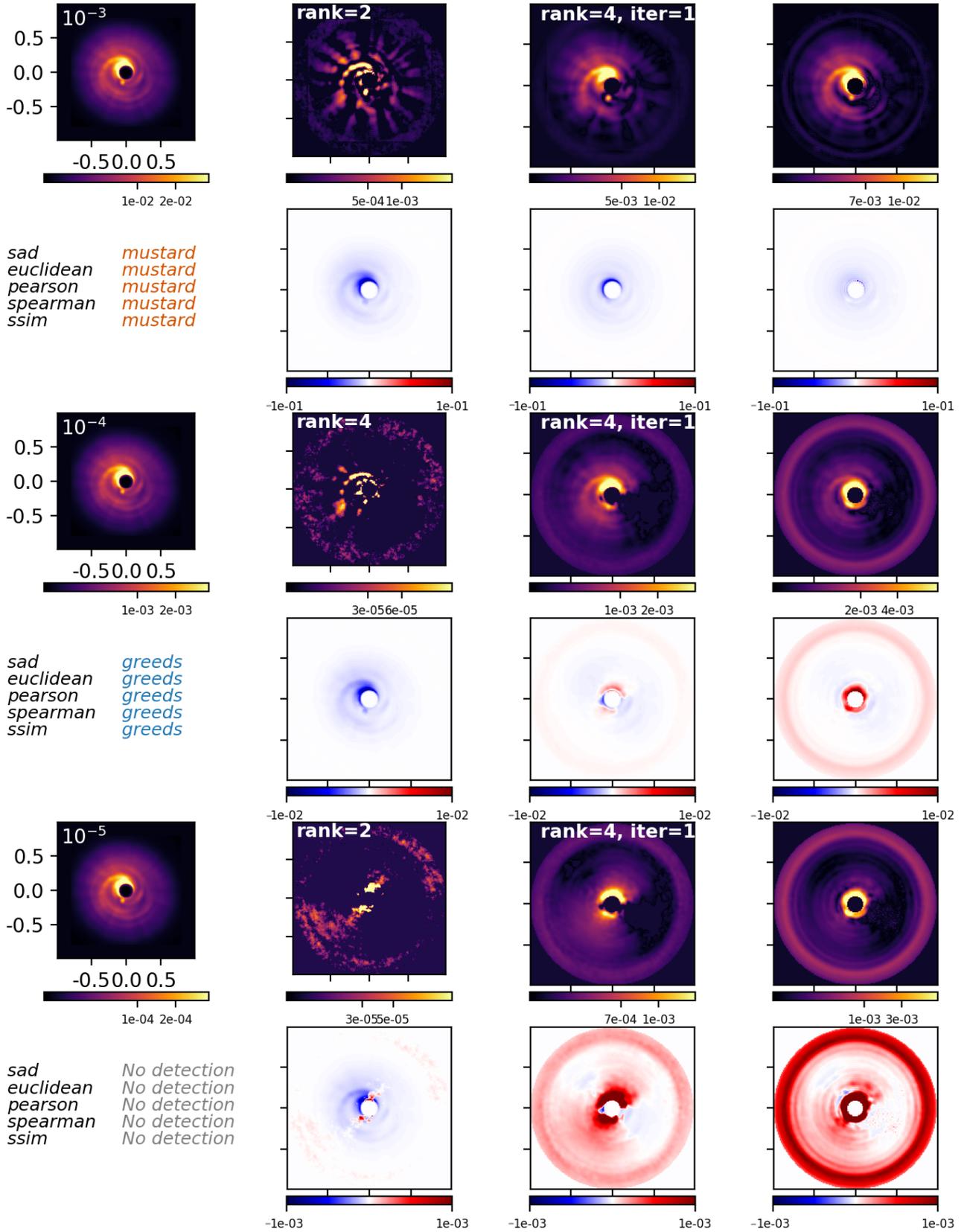}
            \setlength\abovecaptionskip{0cm}
            \caption{Disk estimations obtained with PCA, I-PCA (\texttt{GreeDs}), and \texttt{mustard} for cube \#\cube~and disk \disk. Details on the empty data sets can be found in Table~\ref{tab:datasets} and Sect.~\ref{sec:sample_datasets}. The three pairs of rows correspond to different injected contrast levels: $10^{-3}$ (top), $10^{-4}$ (middle), and $10^{-5}$ (bottom). The top row of each pair shows the estimations, while the bottom row shows the residuals. Each column displays the following: GT (left), PCA (middle-left), I-PCA (middle-right), and \texttt{mustard} (right). The best method, as determined by each metric, is indicated in the bottom-left corner of each pair of rows. More details are provided in the text of Appendix~\ref{sec:simus}.}
            \label{fig:\cube\disk}
        \end{figure*}
        \else
            \begin{figure*}[!ht]
            \centering
            \includegraphics[width=\linewidth]{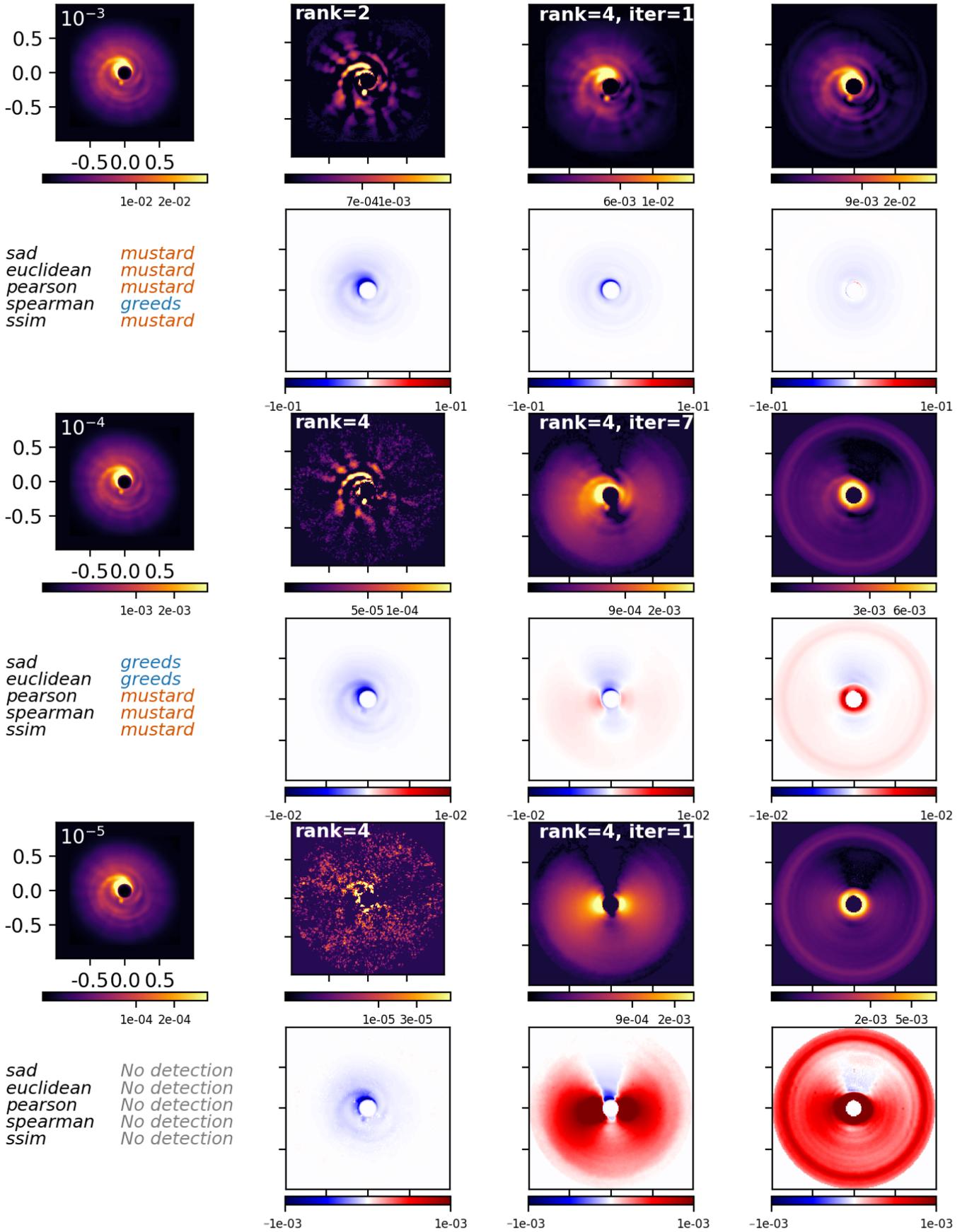}
            \caption{Same as Fig~\ref{fig:4E}, but for the case of cube \#\cube~and disk \disk.}
            \label{fig:\cube\disk}
            
        \end{figure*}
        \fi
    \else
            \begin{figure*}[!ht]
            \centering
            \includegraphics[width=\linewidth]{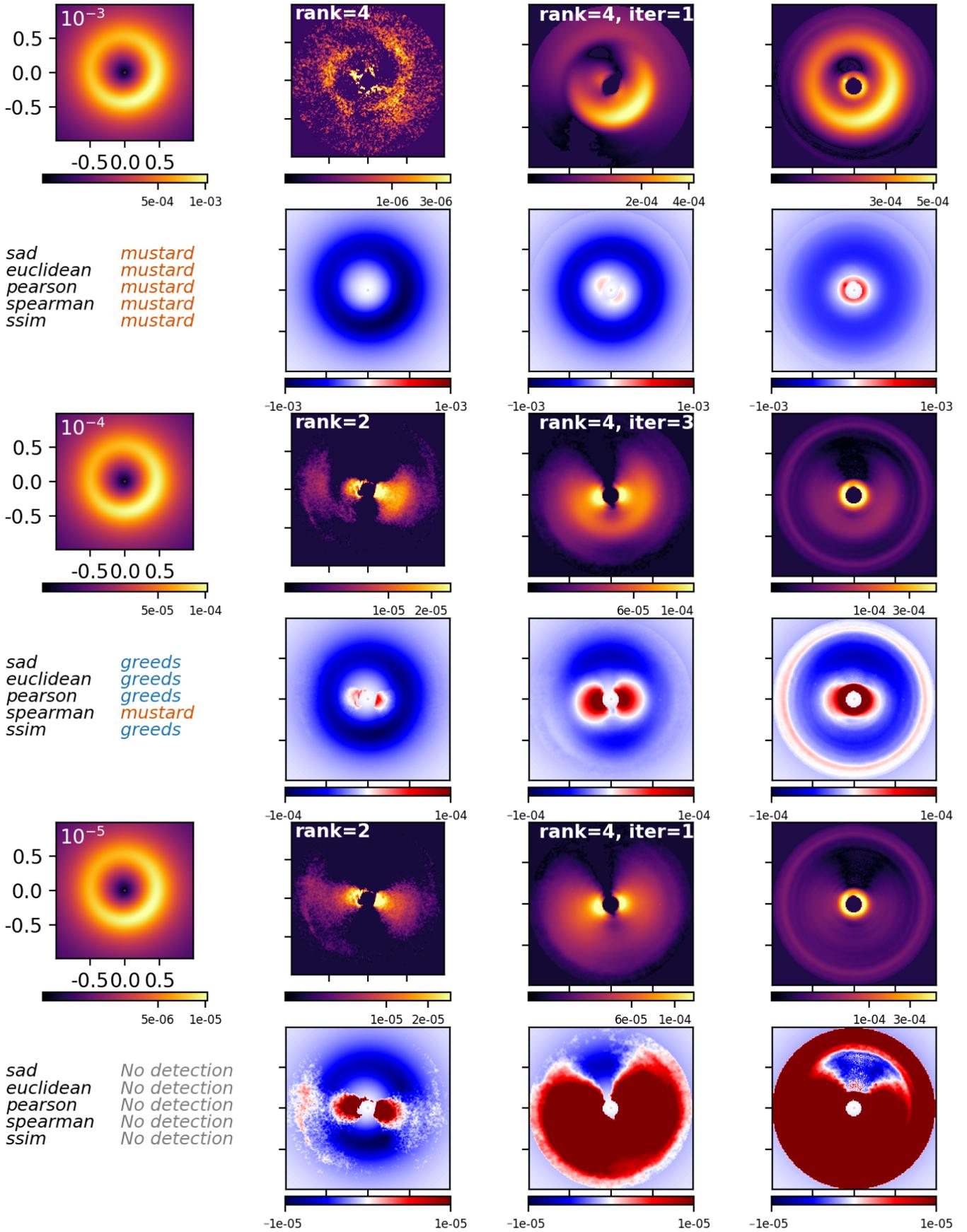}
            \caption{Same as Fig~\ref{fig:4E}, but for the case of cube \#\cube~and disk \disk.}
            \label{fig:\cube\disk}
        \end{figure*}
    \fi
    }

\end{appendix}

\end{document}